\documentclass[10pt]{article}
\usepackage{amstext,amsmath,amssymb}
\usepackage[dvips]{color}

\usepackage{graphicx}

\numberwithin{equation}{section}

\usepackage[active]{srcltx}

\newcommand{\llangle}{\left\langle}
\newcommand{\rrangle}{\right\rangle}
\newcommand{\B}{\vec{B}}
\newcommand{\Bz}{B_{z}}  
\newcommand{\bp}{f}
\newcommand{\bd}{b}
\newcommand{\blo}{\bd_{\|}}
\newcommand{\btr}{\bd_{\perp}}

\newcommand{\bydefl}{:=}
\newcommand{\bydefr}{=:}

\newcommand{\cm}{\,,} 
\newcommand{\lcm}{[}  
\newcommand{\rcm}{]}  

\newcommand{\dm}{\varrho}

\newcommand{\dU}{\Delta U}

\newcommand{\el}{\varepsilon}
\newcommand{\bEm}{\beta\el_{\m}} 
\newcommand{\y}{y}
\newcommand{\ds}{d}

\newcommand{\tf}{\Delta}

\newcommand{\eK}{G}

\newcommand{\half}{\tfrac{1}{2}}
\newcommand{\third}{\tfrac{1}{3}}

\newcommand{\Ham}{{\cal H}}
\newcommand{\Htot}{\Ham}
\newcommand{\Hs}{\Ham_{0}}

\newcommand{\ch}{\,\mathrm{ch}\,}
\newcommand{\sh}{\,\mathrm{sh}\,}
\newcommand{\thrm}{\,\mathrm{th}\,} 
\newcommand{\cth}{\,\mathrm{cth}\,}

\newcommand{\iu}{{\rm i}}
\newcommand{\Id}{\mathbb{I}} %
\newcommand{\kps}{\sigma}
\newcommand{\kpt}{\tau}
\newcommand{\m}{m}
\newcommand{\n}{n}

\newcommand{\K}{D}
\newcommand{\mbra}{\langle \m|}
\newcommand{\mket}{|\m\rangle}
\newcommand{\nbra}{\langle\n|}
\newcommand{\nket}{|\n\rangle}

\newcommand{\lf}{\ell}   

\newcommand{\Mz}{\llangle\m\rrangle}
\newcommand{\MMz}{\llangle\m^{2}\rrangle}

\newcommand{\Ao}{A}
\newcommand{\Bo}{B}

\newcommand{\drm}{\mathrm{d}} 
\newcommand{\e}{\mathrm{e}{}}

\newcommand{\Sm}{S}
\newcommand{\J}{\vec{\Sm}}
\newcommand{\Sx}{\Sm_{x}}
\newcommand{\Sy}{\Sm_{y}}
\newcommand{\Sz}{\Sm_{z}}
\newcommand{\Sco}{\Sm_{0}}
\newcommand{\Scp}{\Sm_{+}}
\newcommand{\Scm}{\Sm_{-}}
\newcommand{\Spm}{\Sm_{\pm}}
\newcommand{\SSp}{\Sm(\Sm+1)}

\newcommand{\Tr}{{\rm Tr}{}}
\newcommand{\kT}{T}

\newcommand{\VQ}{Q}
\newcommand{\V}{V}

\newcommand{\X}{\chi}
\newcommand{\Xlo}{\chi_{\|}}
\newcommand{\Xtr}{\chi_{\perp}}
\newcommand{\Xran}{\overline{\chi}}
\newcommand{\Xc}{\chi_{\rm c}}
\newcommand{\Xising}{\chi_{\rm I}}

\newcommand{\Z}{{\cal Z}}

\begin{document}

\bibliographystyle{ieeetr} 


\title{ \Large equilibrium susceptibilities of superparamagnets: 
longitudinal \& transverse,
quantum \& classical }

\author{\normalsize
J.~L. Garc\'{\i}a-Palacios$^{(1)}$  J.~B. Gong$^{(1,2)}$  and F. Luis$^{(3)}$
\protect\\[1.ex]\parbox{0.9\textwidth}{\centering\scriptsize
$^{(1)}$ 
Department of Physics and Centre of Computational Science \&  Engineering, 
NUS, Singapore 117542
\\
$^{(2)}$ 
NUS Graduate School for Integrative Sciences \& Engineering, Singapore 117597
\\
$^{(3)}$ 
Instituto de Ciencia de Materiales de Arag\'on,
CSIC ---  Universidad de Zaragoza, Spain
}
}

\date{}

\maketitle 

\begin{abstract}
The equilibrium susceptibility of uniaxial paramagnets is studied in a
unified framework which permits to connect traditional results of the
theory of quantum paramagnets, $\Sm=1/2,~1,~3/2$, \dots, with
molecular magnetic clusters, $\Sm\sim5,~10,~20$, all the way up,
$\Sm=30,~50,~100$,~\dots to the theory of classical superparamagnets.
This is done using standard tools of quantum statistical mechanics and
linear response theory (the Kubo correlator formalism).
Several features of the temperature dependence of the susceptibility
curves (crossovers, peaks, deviations from Curie law) are studied and
their scalings with $\Sm$ identified and characterized.
Both the longitudinal and transverse susceptibilities are discussed,
as well as the response of the ensemble with anisotropy axes oriented
at random.
For the latter case a simple approximate formula is derived too, and its
range of validity assessed, so it could be used in modelization of
experiments.
\end{abstract}





\section{introduction}


Modern magnetism, quantum mechanics and statistical mechanics have a
long history of common development since the beginning of the XX
century.
Behind it are the names of the pioneers: Langevin, Brillouin, Bohr
\&~Van Leeuwen, Landau, Van Vleck, Pauli, \dots
After spectroscopy, magnetism helped the most in the understanding of
the atom.
Besides, some of the earliest successes of the Gibbs-Boltzmann distribution
were indeed applications to magnetic problems \cite{white}.
Initially, applications to the simplest magnetic systems, atoms or
small molecules with a permanent magnetic moment: {\em paramagnets}.

When the magnetic moments are brought together they can interact (by
dipole-dipole coupling or by ``exchange'', due to the Pauli
principle).
Then a new game starts, with the possibility of long-range order, spin
waves, etc.
Another important player, when the spins are placed in a molecular
complex or in a solid, is the magnetic anisotropy.
The spin-orbit coupling allows a magnetic moment to sense the electric
field from neighboring ions.
This gives rise to preferred orientations, according to the space
symmetry of the compound/solid: cubic, tetragonal (biaxial), hexagonal
($\sim$uniaxial), etc \cite{chikazumi,aharoni}.

The magnetic anisotropy can also interplay with the spin-spin
interactions, modifying the spin-wave dispersion relations or
determining the spin orientation inside ordered ``domains''
\cite{chikazumi,aharoni}.
If there is no room for domain walls, as in sufficiently fine
particles, the anisotropy would dictate the stable orientations of the
whole magnetic moment (the dipole-dipole interaction also plays a
role, as it ``adds'' to the anisotropy).
Those stable orientations are the basis of using such particles as
physical bits.

\subsection{quantum \& classical superparamagnets}

One more element sets in if the temperature is high enough (but still
below the ordering temperatures), or when the particle is sufficiently
small ($\sim$\,nm).
Then, by thermal activation, the net moment can overcome the
anisotropy energy barriers and flip back and forth between the minima
\cite{nee49,panpol93}.
This is a nightmare for magnetic recording \cite{white2001}, but a
blessing for new phenomenology.
Inside we have an ordered magnet, which seen from the outside
resembles a paramagnet, but with a very large magnetic moment,
$\Sm\sim100$--$1000$ --- {\em superparamagnetism}.
And yet another connection with statistical mechanics is established,
as the non-equilibrium orientational distribution of those
``classical'' spins is described with the tools of Brownian/stochastic
dynamics, e.g., Fokker--Planck or Langevin equations
\cite{bro63,kubhas70,gar2000}.

The 80s and 90s brought, through the advances in chemical synthesis, a
new member of the paramagnetic family, {\em molecular magnetic
clusters\/} \cite{blupra2004} (also refereed too as {\em
single-molecule magnets}).
These are made of complex molecules with a net spin
$\Sm\sim5,~10,~20$, somewhere in between the traditional paramagnets
and the superparamagnets.
But, contrary to magnetic nanoparticles, they are assembled in
molecular crystals, minimizing extrinsic sources of dispersion (in
size, anisotropy parameters, etc.) and turning the comparison between
experiment and theory cleaner.

As for the static part of the magnetic anisotropy, it can be quite
large in molecular magnets, so the stable orientations could provide
the basis of a quantum bit.
Nevertheless, environmental effects (decoherence in particular),
emerge already from their own nuclear spins or from the dynamical part
of the spin-orbit coupling (neighboring ions are never static; the
lattice oscillations modulate the local electric field, which modifies
the orbital motions, eventually affecting $\J$).
Again, a curse for technology, but a blessing for physics
\cite{leg2002}.
The total system can be described by a spin-phonon Hamiltonian
\cite{pake,harpolvil96,garchu97}, and constitutes a neat realization
of the spin-boson paradigm of the theory of quantum dissipative
systems \cite{weiss}; but with a $\Sm\geq1/2$ immersed in a bath of
true ``harmonic oscillators'' (the lattice phonons).

\subsection{the spin Hamiltonian}

The theoretical description of paramagnets starts with a spin
Hamiltonian, where one replaces the true microscopic Hamiltonian (or a
part of it) by one written in terms of spin energy levels and
operators (a procedure initiated by Heisenberg, Dirac \& Van Vleck
\cite[\S~72]{lanlif9}).
This provides an equivalent description of the statics (based on the
spectrum) and of part of the dynamics.
The inner complexities, symmetrization of wave functions, local
fields, spin-orbit coupling, are left in the backstage and we only see
their effect through the spin Hamiltonian \cite{white,pake,lanlif9}.

\begin{figure}[!t]
\centerline{
\includegraphics[width=6.5cm]{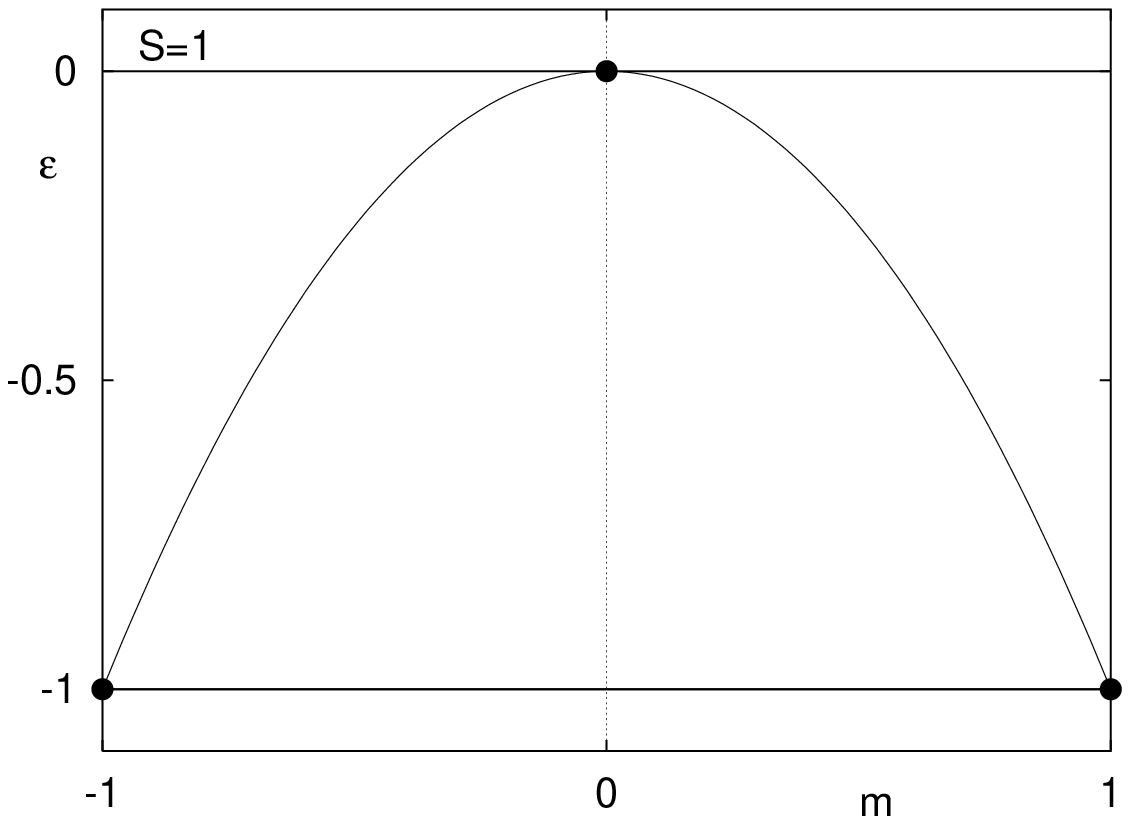}
\hspace*{-2.ex}
\includegraphics[width=6.5cm]{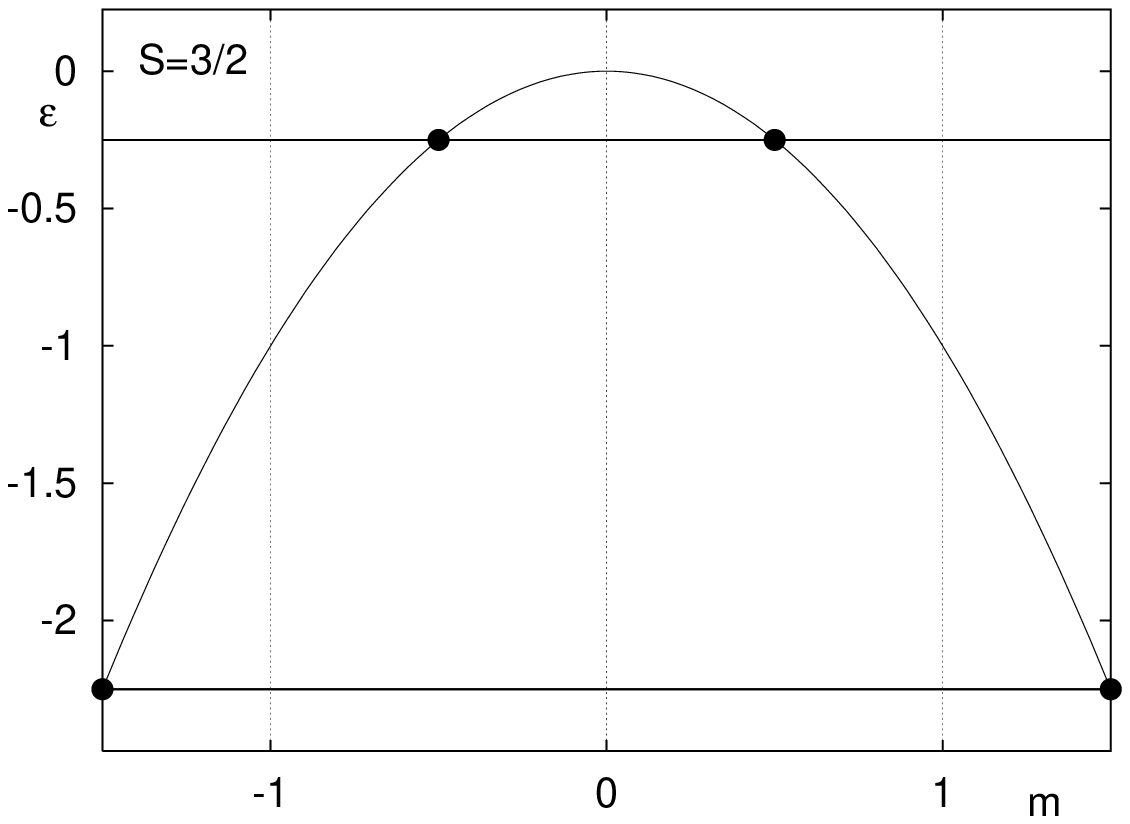}
}
\centerline{
\includegraphics[width=6.5cm]{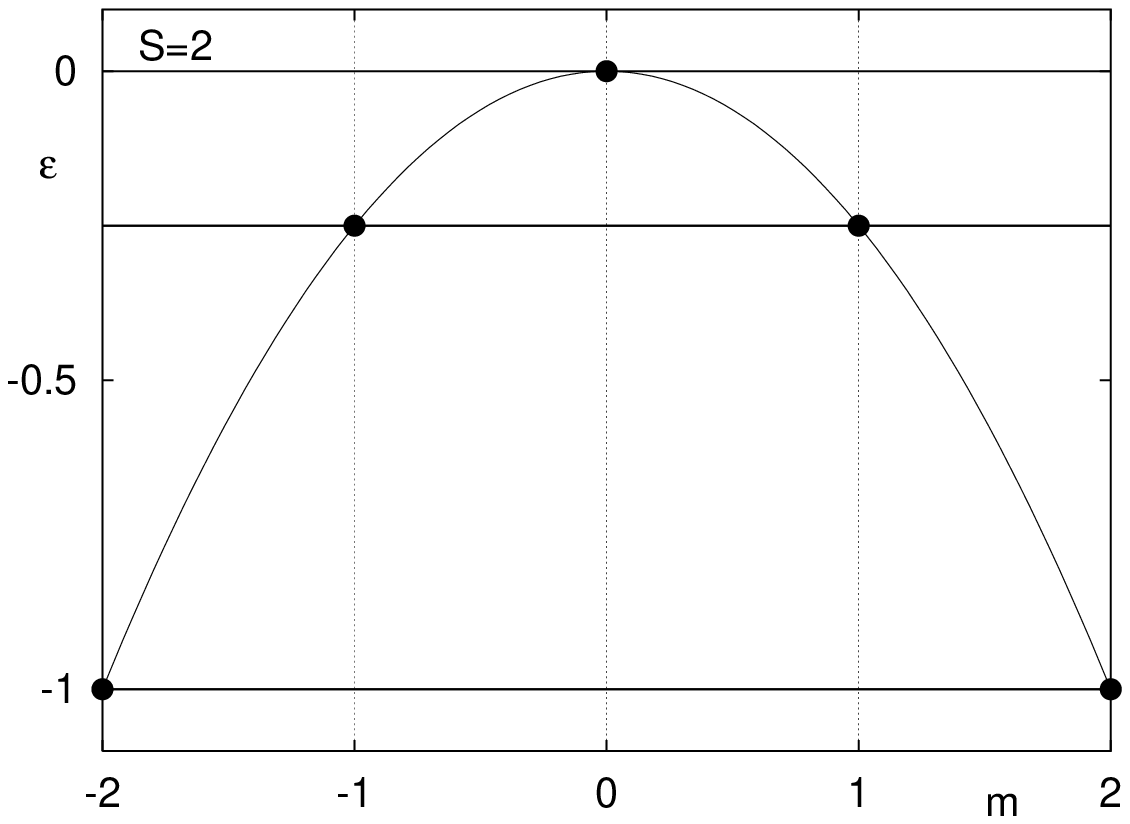}
\hspace*{-2.ex}
\includegraphics[width=6.5cm]{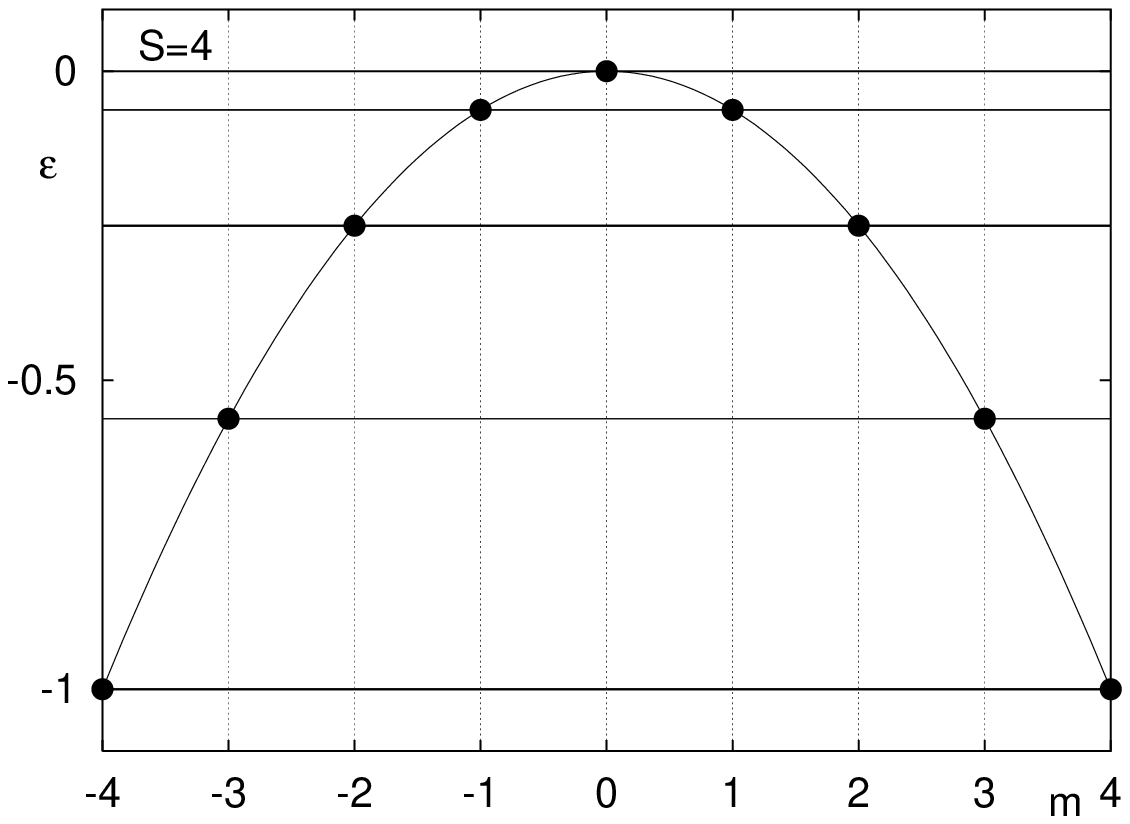}
}
\caption{
Energy levels $\el_{\m}=-\K\m^{2}-\Bz\m$ of anisotropic spins
$\Sm=1$, $3/2$, $2$, and $4$ at zero field.
For $\Sm=1$ and $3/2$ we set $\K=1$.
For $\Sm=2$ and $\Sm=4$, the anisotropy constant is scaled as
$\K=1/\Sm^{2}$, so that these spins have the same energy barrier as
$\Sm=1$.
}
\label{fig:levels}
\end{figure}

For many classical superparamagnets \cite{upasrimeh2000} and molecular
clusters \cite{blupra2004}, a good first approximation is provided by
%
\begin{equation}
\label{Huni:B}
\Hs
=
-\K\,\Sz^{2}-\B\cdot\J 
\;,
\end{equation}
with $\K$ the uniaxial anisotropy constant.
$\K>0$ produces two preferred spin orientations (the ``bit'',
Fig.~\ref{fig:levels}) and brings in the physics of bistable systems
(e.g., thermal activation over a potential barrier).
But there is also a rich ``coherent'' or transverse dynamics
\cite{pake,garzue2006}, with spin resonances, absorption peaks, etc.%
\footnote{
The big spin Hamiltonian~(\ref{Huni:B}) is also an effective
description of a collection of $N$ two-level systems (not necessarily
magnetic), used in atom optics since the 70s \cite{ACGT72}.
%
%
Non-interacting entities correspond to $\K=0$, while $\K\neq0$ with a
transverse field $B_{x}$ describes certain types of uniform
interaction (the Lipkin-Meshkov-Glick model originally from nuclear
physics).
The theoretical work on this front was revived with the boom of cold
gases and condensates \cite{miletal97,angvar2001}.
} 

\subsection{the magnetic susceptibility \& linear-response theory}
\label{LRT}

Most of what we know about magnets, and even their names and the ways
we classify them, is based on their characteristic response
\cite{white}.
That is, on how they respond to the application of controlled
magnetic field probes.
When the probes are small enough, linearizations and perturbation theory,
in its different guises, can be used by the theorist.

In molecular magnetism a procedure due to Van Vleck
\cite[Ch.~3]{white} became quite popular.
The energy levels of a paramagnet, in the presence of the probing
field, are obtained using quantum-mechanical perturbation theory, and
the result plugged into a Gibbs-Boltzmann formula.
In other, more complicated problems of theoretical magnetism (e.g.,
quantum spin chains
\cite{thotho1983,chatterjee1986,jarloh1986,minami96} or inner
structure of molecular clusters \cite{zvedobharkat98}) other tools
from linear response theory are employed.
The Kubo correlator, in particular, provides a formal general solution
for the response functions, the susceptibility $\X$, without invoking
perturbed levels at all (the usual ``magic'' of linear-response theory
\cite{kuboII,dattagupta}).

\subsection{our approach and goals}

We aim to calculate the equilibrium linear response for paramagnets
with simple anisotropies.
But in a way that could connect the traditional results for quantum
paramagnets, all the way up to the results of classical
superparamagnetic theory.

The approach of Van Vleck, at least in its popular form of diagonalize
\&~linearize \cite{carlin,kahn}, becomes awkward for moderately small
$\Sm$.
Therefore we will resort to the Kubo correlator formalism, and
particularize it with the (super)paramagnetic problem in mind.

Schematically, the approach takes two steps (Sec.~\ref{derivX}):
(1) Obtain the correlator
$\int_{0}^{\beta}\!
\drm\kps\,
\Tr\,\big[
\e^{(\beta-\kps)\Hs}\,
\V\,
\e^{+\kps\Hs}\,
\V
\big]$,
and sandwich here and there identities $\Id=\sum_{\m}\mket\mbra$
formed with the unperturbed levels $\Hs\mket=\el_{\m}\mket$.
This gives a susceptibility
$\X
\sim
\beta
\sum_{\m\,\n}
\e^{-\bEm}\,
|\V_{\m\,\n}|^{2}
K[\beta(\el_{\m}-\el_{\n})]$
with the Kubo function $K(X)=(\e^{X}-1)/X$ evaluated at the level
differences.
(2) Next, for uniaxial problems, $\Hs=-\K\,\Sz^{2}-\Bz\,\Sz$, the
    matrix elements of the perturbation $\V$ simplify,
    $\V_{\m\,\n}\sim\V_{\|}\delta_{\m,\n}+\V_{\perp}\delta_{\m,\n\pm1}$,
    and the above $\X$ can be reduced to a single sum over the
    unperturbed spectrum (in the line of naive linear-response
    theory).
One component of $\X$, the longitudinal susceptibility, comes from the part
of $\V$ commuting with $\Hs$; the non-commuting part gives the
transverse susceptibility [Eqs.~(\ref{Xlo})--(\ref{Xtr})].

Some advantages of this approach are:
(1) The intermediate step of getting eigenvalues in presence of the
    perturbation is bypassed.
(2) There is no small-denominators problem; $K(0)=1$ ensures
    automatically the finiteness of the perturbative result, without
    the need of manual pairing of degenerate levels or the like
    \cite{serber}.
(3) The susceptibility formula, expressed as a {\em single\/} sum over
    the spin levels $\m$, is ready to use for any $\Sm$.

As a check, we recover in Sec.~\ref{sec:PM} a number of traditional
results for paramagnets, including the transverse response of small
spins $\Sm=1$ and $\Sm=3/2$.
Then, we will move on to arbitrary $\Sm$ in Sec.~\ref{sec:SPM} and
discuss some features of the temperature-dependent susceptibility
curves, like crossovers, peaks, and the isotropy of $\X$ in the
randomly oriented ensemble.
We will study how these features evolve with $\Sm$ under appropriate
scalings (Fig.~\ref{fig:levels}).

The interest in the $\X(T)$ curves stems from their routine use as
indicators of anisotropies, interactions between spins,
etc. \cite{carlin,kahn,maurer}.
Therefore it is important to understand well the paramagnetic ``ideal
gas'' (as the implicit reference curve in those procedures), including
the effect of simple anisotropies (e.g., uniaxial, biaxial, etc.).
We hope that our results will convince the reader of the advantage of
having a unified framework for the response of the whole paramagnetic
family, from the traditional members, then molecular magnetic
clusters, up to classical superparamagnets.


\section{derivation of the susceptibility}
\label{derivX}

Instead of starting writing down directly the linear susceptibility in
terms of a Kubo correlator, we will briefly discuss its basis on plain
statistical-mechanical perturbation theory.
This will make the presentation more self-contained, and will hint at
how to proceed if higher orders are required, beyond linear response.


\subsection{statistical mechanical preliminaries}

Statistical mechanics is about computing averages,
variances/dispersions, etc.\ for a given system with the sole input of
the {\em system Hamiltonian\/} $\Htot$ and a few parameters
incorporating the effect/presence of a thermalizing environment (like
$\beta=1/\kT$, chemical potentials, etc.).
%
%
%
The averages are generated from the {\em density matrix\/}
$\dm\propto\e^{-\beta\Htot}$, and for a given operator quantity $\VQ$
one forms the following basis-independent object
%
\begin{equation}
\label{average}
\llangle \VQ \rrangle
=
\Tr\,
\big(
\dm\,
\VQ
\big)
=
\sum_{\m\n}
\dm_{\m\n}
\VQ_{\n\m}
\;.
\end{equation}
For consistency of the definition, $\dm$ is ``normalized'' to have
unit trace $\Tr\,(\dm)=1$ (so the average of a constant is the
constant itself):
%
\begin{equation}
\label{dm:Z}
\dm
=
\e^{-\beta\Htot}
/
\Z
\;,
\qquad
\Z
=
\Tr\,
\e^{-\beta\Htot}
\;.
\end{equation}
The normalization function, $\Z$, plays a central role in several
parts of the formalism.
In particular, as an auxiliary/abridged generator of averages (by
differentiation) with the advantage of being a scalar instead of an
operator.


\subsubsection{response observables}
\label{avg:response}


Let us imagine that the quantity we are interested in can be obtained
from the Hamiltonian by differentiation
%
\begin{equation}
\label{VQfromHam}
\VQ
=
-
\partial_{\bp}
\Htot
\qquad
(\mbox{operator identity})
\;.
\end{equation}
Say, the total Hamiltonian is of the form
$\Htot=\Hs-\bp\,\VQ$,
as in the examples of force--coordinate, $-F\,x$, or
field--dipole/spin, $-\B\cdot\J$.
Then the average of $\VQ$ can be obtained from the partition function
as follows
%
\begin{equation}
\label{VQ:average}
\llangle \VQ \rrangle
=
\frac{1}{\Z}
\frac{1}{\beta}
\Tr\,
\big(
\partial_{\bp}\,
\e^{-\beta\Htot}
\big)
=
\frac{1}{\Z}
\partial_{(\beta\bp)}\,
\Tr\,
\big(
\e^{-\beta\Htot}
\big)
=
\frac{\Z'}{\Z}
\;.
\end{equation}
Here $(\:)'$ denotes derivative with respect to the ``thermal''
force/field parameter $\xi\bydefl\beta\bp$ entering in
$-\beta\Htot=-\beta\Hs+\xi\,\VQ$.~%
\footnote{
The proof of (\ref{VQ:average}) is a bit less direct than it looks
like, because one cannot differentiate the exponential of a matrix
plainly as
$\partial_{\xi}\,\e^{\Ao(\xi)}\neq\e^{\Ao(\xi)}\partial_{\xi}\Ao$,
when $\Ao$ and $\partial_{\xi}\Ao$ do not commute.
However, using the proper identity
$\partial_{\xi}\,\e^{\Ao(\xi)}
=
\int_{0}^{1}\!
\drm\kps\,
\e^{(1-\kps)\Ao(\xi)}\,
(\partial_{\xi}\Ao)
\e^{+\kps\Ao(\xi)}$,
with $\kps$ an auxiliary parameter, along with the cyclic property of
the trace $\Tr\,(C\,D)=\Tr\,(D\,C)$, one arrives at the desired
result:
$\partial_{\xi}\Tr\,\e^{\Ao(\xi)}
=
\Tr\,\big(\e^{\Ao}\partial_{\xi}\Ao\big)$,
with the trace removing any ordering problem.
} 

We will refer to the average of a quantity $\VQ$ derivable from the
Hamiltonian as above as a {\em response observable}.
This definition does not include pure thermal quantities, like the
thermodynamical energy or the specific heat, although these can be
obtained by differentiating $\Z$ with respect to $\beta=1/\kT$.


\subsubsection{susceptibility \& derivatives of $\Z$}

As the force/field parameter $\bp$ is at our disposal, we can change
$\bp$ and from the induced change in $\llangle \VQ \rrangle$ learn how
the system ``responds''.
A natural quantity then arises
%
\begin{equation}
\label{X:def}
\X
=
\partial_{\bp}
\llangle \VQ \rrangle
\;,
\end{equation}
quantifying the sensitivity of the system to changes in $\bp$, and for
this reason called the {\em susceptibility}.
$\X$ is thus a second derivative of $\Z$, and if we keep on using
$\xi=\beta\bp$ and $(\:)'=\partial_{\xi}$, we have
$\X = \beta [ (\Z''/\Z) - (\Z'/\Z)^{2} ]$,
where the square brackets is merely $(\Z'/\Z)'$.

We can think that $\bp$ is the part of the force/field that
we change (the probe), over a fixed bias $\bp_{0}$.
Then, from $-(\bp_{0}+\bp)\,\VQ$ we can move $-\bp_{0}\,\VQ$ into the
unperturbed part $\Hs$, and eventually evaluate our derivatives at
$\bp=0$ (linear susceptibility).
Then, all we need to do is to obtain the $\xi$-expansion of $\Z$ to
second order, $\Z\simeq\Z_{0}+\xi\Z_{1}+\half\xi^{2}\Z_{2}$, and the
linear susceptibility
%
%
will simply follow as a combination of the expansion coefficients
%
\begin{equation}
\label{X:Zi}
\X
=
\beta
\bigg[
\frac{\Z''}{\Z}
-
\Big(
\frac{\Z'}{\Z}
\Big)^{2}
\bigg]\bigg|_{\xi=0}
=
\beta
\bigg[
\frac{\Z_{2}}{\Z_{0}}
-
\Big(
\frac{\Z_{1}}{\Z_{0}}
\Big)^{2}
\bigg]
\;.
\end{equation}


\subsection{
perturbative treatment
}
\label{subsec:pertheo}

When one proceeds to expand the partition function
%
\begin{equation}
\Z
=
\Tr\,
\e^{-\beta(\Hs+\V)}
\;,
\qquad
\V=-\bp\,\VQ
\;,
\end{equation}
with respect to $\xi=\beta\bp$, one faces the problem of handling
exponentials of operators or matrices.
Classically, one proceeds by factorizing and expanding 
$\e^{a+b}=\e^{a}\,\e^{b}\simeq\e^{a}(1+b+b^{2}/2+\cdots)$.
For operators, however, plain factorization does not hold,
and one has instead a Baker-Campbell-Hausdorff formula
%
\begin{equation}
\label{BCH}
\e^{\Ao+\Bo}
=
\e^{\Ao}\,
\e^{\Bo}\,
\e^{-\frac{1}{2}\lcm\Ao\cm\Bo\rcm}\,
\e^{+\frac{1}{3}
\left(
\lcm\Ao\cm\lcm\Ao\cm\Bo\rcm\rcm
+
\lcm\Bo\cm\lcm\Ao\cm\Bo\rcm\rcm
\right)}\,
\cdots
\;.
\end{equation}
This kind of expression is useful if one can recurrently write higher
order commutators, e.g. $\lcm\Ao\cm\lcm\Ao\cm\Bo\rcm\rcm$, in terms of
lower order ones, $\Ao$, $\Bo$, $\lcm\Ao\cm\Bo\rcm$.
For example, when $\Ao\sim\Sz$ and $\Bo\sim\Spm=\Sx\pm\iu\Sy$
(isotropic spin), or when $\Ao\sim b_{+}b_{-}$ and $\Bo\sim b_{\pm}$,
with $[b_{-},b_{+}]=\Id$ (harmonic oscillators).
But in the general case equation~(\ref{BCH}) is of little use to do
perturbation theory in one of the operators (actually, already for
$\Ao\sim\Sz^{2}$ and $\Bo\sim\Spm$).


\subsubsection{
perturbations from a Kubo identity  (interaction picture)
}

A way out in the problem of handling operator exponentials is provided
by Kubo identities of the type \cite[p.~148]{kuboII}
%
\begin{equation}
\label{kuboid2}
\e^{\Ao+\Bo}
=
\e^{\Ao}\,
\Big[
\Id
+
\int_{0}^{1}\!
\drm\kps\,
\e^{-\kps\Ao}\,
\Bo\,
\e^{+\kps(\Ao+\Bo)}\,
\Big]
\;,
\end{equation}
where one resorts to an integral over an auxiliary parameter.%
\footnote{
The same kind of auxiliary parameter, with in principle no physical
meaning, we used to handle $\partial_{\xi}\,\e^{\Ao(\xi)}$.
The proof of Eq.~(\ref{kuboid2}) follows in four simple steps:
(i) Isolate the integral by left multiplying by $\e^{-\Ao}$.
(ii) Define
    $F(\lambda)\bydefl\e^{-\lambda\,\Ao}\,\e^{+\lambda(\Ao+\Bo)}$; then
    $F(0)=\Id$, while $F(1)$ is the target.
(iii) Differentiate with respect to $\lambda$ (no ordering problems) to
    get $F'(\lambda)=\e^{-\lambda\,\Ao}\,\Bo\,\e^{\lambda(\Ao+\Bo)}$;
    we are almost done [cf.~the integrand in Eq.~(\ref{kuboid2})].
(iv) Integral reconstruction
    $F(1)=F(0)+\int_{0}^{1}\!\drm\kps\,F'(\kps)$ gives the
    $\big[\Id+\int_{0}^{1}\cdots\big]$ on the right-hand side
    of~(\ref{kuboid2}).
} 
The integrand includes $\Bo$ (now freed) and again the exponential of
the sum.
Therefore, by iterating the same expression, 
$\e^{\kps(\Ao+\Bo)}
=
\e^{\kps\Ao}\,
[
\Id
+
\int_{0}^{\kps}\!
\drm\kpt\,
\e^{-\kpt\Ao}\,
\Bo\,
\e^{+\kpt(\Ao+\Bo)}
]$
one generates the successive powers of $\Bo$ and can do perturbation
theory.

For example, to get second-order derivatives of $\Z$, one just needs to
iterate to second order.
Then the $\Bo$ in the last $\e^{+\kpt(\Ao+\Bo)}$ is dropped (as there
are already two $\Bo$ multiplying it), and one is left with
%
\begin{eqnarray}
\label{kuboid2:iterate2}
\e^{\Ao+\Bo}
\simeq
\e^{\Ao}\,
\Big[
\;
\Id
&+&
\int_{0}^{1}\!
\drm\kps\,
\overbrace{
\e^{-\kps\Ao}\,
\Bo\,
\e^{+\kps\Ao}\,
}_{}
\nonumber\\
&+&
\int_{0}^{1}\!
\drm\kps\,
\e^{-\kps\Ao}\,
\Bo\,
\e^{+\kps\Ao}\,
\int_{0}^{\kps}\!
\drm\kpt\,
\e^{-\kpt\Ao}\,
\Bo\,
\e^{+\kpt\Ao}
\Big]
\;.
\end{eqnarray}
Here we recognize the ``interaction-picture'' evolution
$\e^{-\kps\Ao}\,\Bo\,\e^{+\kps\Ao}$.
Indeed the structure of the above formula is present in most of
quantum mechanical perturbation theory \cite{merzbacher}, including
scattering; it is also used to derive weak-coupling master equations
in quantum open systems \cite{kuboII,zueco}.


\subsubsection{
tracing \& the Kubo correlator
}
\label{tracing}

Tracing is now required to get $\Z=\Tr\,\e^{-\beta\Htot}$.
The trace will simplify the perturbative treatment (as compared with
perturbed time evolutions), by allowing to move operators around in
combinations like $\e^{-\kps\Ao}\,\Bo\,\e^{+\kps\Ao}$, using the
trace's cyclic property.
We will undertake this first, arriving at the structure of the {\em
Kubo correlator} \cite{morkat69}, and then simplify further by doing
the trace in the eigenbasis of $\Ao$.


\paragraph{
simplifying the trace by cycling.
}
Using $\Tr\,(C\,D)=\Tr\,(D\,C)$ when tracing the second term in
(\ref{kuboid2:iterate2}) one gets rid of the integral, as the first
exponential can be moved to the end, canceling the dependence on the
auxiliary variable $\kps$.
The third term also simplifies, following a procedure explained in
most field-theory books.%
\footnote{
See for example \cite[Ch.~8.3]{greiner-FQ}.
One first converts the integral over the triangle $\kps\in[0,1]$
$\kpt\in[0,\kps]$ into the integral over the unit square
$[0,1]\times[0,1]$, by using ``chronological'' ordering (this yields a
factor $1/2$).
Then the cyclic property of the trace is used to show that the
integrand does not depend on one of the integration parameters.
} 
Collecting the results one arrives at
%
\begin{equation}
\label{kuboid2:iterate2:trace}
\Tr\,\big(\e^{\Ao+\Bo}\big)
\simeq
\Tr\,\big(\e^{\Ao}\big)
+
\Tr\,\big(\e^{\Ao}\,\Bo\big)
+
\half
\int_{0}^{1}\!
\drm\kps\,
\Tr\,\big[
\e^{(1-\kps)\Ao}\,
\Bo\,
\e^{+\kps\Ao}\,
\Bo\,
\big]
\;.
\end{equation}
We see that one of the auxiliary parameters, $\kps$, is still with us; indeed
the last term is a bare form of the Kubo correlator,
$\int_{0}^{\beta}\!
\drm\kps\,
\Tr\,\big[
\e^{(\beta-\kps)\Hs}\,
\V\,
\e^{+\kps\Hs}\,
\V
\big]$.

The goal is therefore accomplished: tracing the exponential of a sum
of operators, to second order in one of them.
For this Baker-Campbell-Hausdorff is not suited, while a classical
handling of the operators
$\e^{\Ao+\Bo}\simeq\e^{\Ao}(1+\Bo+\Bo^{2}/2)$,
would give correct results only to first order.


\paragraph{
tracing in the unperturbed eigenbasis.
}
When the eigenstructure of the operator $\Ao$ is known,
$\Ao\mket=a_{\m}\mket$, the trace can be written explicitly in terms
of the eigenvalues $a_{\m}$ and the matrix elements of the
perturbation
%
\begin{equation}
\label{eigenstructure:A}
\Ao\mket
=
a_{\m}\mket
\qquad\quad
B_{\m\,\n}
=
\mbra\Bo\nket
\;.
\end{equation}
Indeed sandwiching identities $\Id=\sum_{\n}\nket\nbra$ between
$\e^{\Ao}$ and $\Bo$, the three parts of
Eq.~(\ref{kuboid2:iterate2:trace}) give
%
\begin{eqnarray}
\label{kuboid2:iterate2:trace:expl}
& &
\Tr\,\big(\e^{\Ao}\big)
=
\sum_{\m}
\e^{a_{\m}}
\;,
\qquad
\Tr\,\big(\e^{\Ao}\,\Bo\big)
=
\sum_{\m}
\e^{a_{\m}}\,
B_{\m\,\m}
\\
& &
\half
\int_{0}^{1}\!
\drm\kps\,
\Tr\,\big[
\e^{(1-\kps)\Ao}\,
\Bo\,
\e^{+\kps\Ao}\,
\Bo\,
\big]
=
\half
\sum_{\m\,\n}
\e^{a_{\m}}\,
|B_{\m\,\n}|^{2}
K(a_{\n}-a_{\m})
\nonumber
\;,
\end{eqnarray}
where we have written $B_{\m\,\n}B_{\n\,\m}=|B_{\m\,\n}|^{2}$
for Hermitian $\Bo$.
The integral over $\kps$ produced the Kubo function
%
\begin{equation}
\label{kubofunction}
K(X)
\bydefl
\int_{0}^{1}\!
\drm\kps\,
\e^{\kps\,X}\,
=
\big(\e^{X}-1\big)/X
\;,
\end{equation}
which enters evaluated at all eigenvalue differences
$X=a_{\n}-a_{\m}$.
This function will follow us all the way to the final expressions.%
\footnote{
Note the connection $K(X)=1/W_{1}(X)$, with
$W_{k}(X)=X^{k}/(\e^{X}-1)$ a transition rate from the theory of open
quantum systems ($k=1$ corresponds to an ``Ohmic'' bath
\cite{weiss,zueco}).
Thus, the Kubo function enjoys a ``detailed balance'' relation as well:
$K(-X)=\e^{-X}\,K(X)$, useful in some manipulations.
} 
%


\subsubsection{
application to the original perturbative problem
}

Let us write down the explicit correspondence with our original
perturbative trace problem $\Z=\Tr\,\exp[-\beta(\Hs-\bp\,\VQ)]$
%
\begin{equation}
\label{}
-\beta\Htot
=
\underbrace{-\beta\Hs}_{\Ao}
+
\underbrace{\overbrace{\xi}^{\beta\bp}\,\VQ}_{\Bo}
\;.
\end{equation}
Now the unperturbed eigenstructure reads
%
\begin{equation}
\label{eigenstructure:Hs}
\Hs\mket
=
\el_{\m}\mket
\quad
\leadsto
\quad
a_{\m}
=
-\bEm
\qquad
B_{\m\,\n}
=
\xi\,
\VQ_{\m\,\n}
\;.
\end{equation}
Then comparison of
Eqs.~(\ref{kuboid2:iterate2:trace}) and~(\ref{kuboid2:iterate2:trace:expl})
with the $\xi$-expansion of
$\Z\simeq\Z_{0}+\xi\Z_{1}+\half\xi^{2}\Z_{2}$, 
gives the sought for coefficients
%
\begin{equation}
\label{Z:pert:expl}
\Z_{0}
=
\sum_{\m}
\e^{-\bEm}
\;,
\quad
\Z_{1}
=
\sum_{\m}
\e^{-\bEm}\,
\VQ_{\m\,\m}
\;,
\quad
\Z_{2}
=
\sum_{\m\,\n}
\e^{-\bEm}\,
|\VQ_{\m\,\n}|^{2}
K_{\m\,\n}
\;.
\end{equation}
In the last term we have introduced the shorthand
%
\begin{equation}
\label{kubofunction:mn}
K_{\m\,\n}
\bydefl
K[\beta(\el_{\m}-\el_{\n})]
\bydefr
K(\beta\tf_{\m\,\n})
\;,
\end{equation}
with the Kubo function evaluated at the level differences
$\tf_{\m\,\n}\bydefl\el_{\m}-\el_{\n}$ (``transition'' frequencies).


\paragraph{
absence of small denominators problems.
}
A final remark on perturbation theory with close levels.
If we combine the denominator $\el_{\m}-\el_{\n}$ from $K_{\m\,\n}$
with the matrix element $|\VQ_{\m\,\n}|^{2}$, one actually sees the
structure of plain quantum-mechanical perturbation theory:
$\VQ_{\m\,\n}\VQ_{\n\,\m}/(\el_{\m}-\el_{\n})$.
However, there is no need to handle degenerate levels, if existing, in
a special way (i.e., no need of pairing, etc. \cite{serber}).
The formalism ensures that the {\em finite temperature\/} perturbative
treatment is finite as well.
Indeed, Taylor expansion of $K(X)=(\e^{X}-1)/X$
%
\begin{equation}
\label{kubofunction:taylor}
K(X)
\simeq
1
+
X/2
+
X^{2}/6
+
\cdots
\;,
\end{equation}
shows that degenerate levels $\tf_{\m\,\n}=0$ would contribute a
finite $K(0)=1$.
This property spares us with a degenerate perturbation theory to
handle close levels $\el_{\m}\simeq\el_{\n}$, and is built in the
finite-temperature formalism (to second order at least).


\subsection{
formulas for the susceptibility
}
\label{subsec:susceptibility}

Now it is immediate to write down explicit expressions for the
susceptibility, by plugging the $\Z$ expansion
coefficients~(\ref{Z:pert:expl}) into 
$\X
=
\beta
\big[
(\Z_{2}/\Z_{0})
-
(\Z_{1}/\Z_{0})^{2}
\big]$.
This form is generic for what we called ``response observables'' in
Sec.~\ref{avg:response}, that is
$\X=\partial_{\bp} \llangle \VQ \rrangle$
with $\VQ=-\partial_{\bp}\Htot$, as in the examples of a coordinate,
$-F\,x$, a spin component, $-\Bz\Sz$, etc.

In what follows we will address the structure of the resulting
susceptibilities, and then particularize the discussion to ``ladder
perturbations'', which include simple mechanical oscillators and
uniaxial paramagnets.


\subsubsection{general $\X$ at zero bias}

As a quick illustration, in the unbiased case where $F=0$ or $\Bz=0$,
one has $\Z_{1}=0$ as well, and the susceptibility simply reads
(we will restore $\Z_{1}\neq0$ shortly)
%
\begin{equation}
\label{X:unbiased}
\X
=
\frac{\beta}{\Z_{0}}
\sum_{\m\,\n}
\e^{-\bEm}\,
|\VQ_{\m\,\n}|^{2}
K_{\m\,\n}
\;.
\end{equation}
This is the kind of ready-to-use expression we mentioned in the
introduction. 
It is written fully in terms of the unperturbed eigenstructure
$\{\el_{\m},\mket\}$, $\VQ_{\m\,\n}=\mbra\VQ\nket$, and
$K_{\m\,\n}=K[\beta(\el_{\m}-\el_{\n})]$.
It looks like a average over the unperturbed system, resembling the
classical result $\chi\sim\langle\Sm_{i}^{2}\rangle$, but with
$K_{\m\,\n}$ encoding effects of non-commutativity of the perturbation
$\VQ$ and the base Hamiltonian. 
With simple numerical diagonalization, the $\X$ above can be used for
a non-linear oscillator,
$\Hs\sim-\half k\,x^{2}+\tfrac{1}{4}q\,x^{4}$,
or arbitrary anisotropic spins,
$\Hs\sim-\K\,\Sm_{z,\pm}^{2}-K\Sm_{z,\pm}^{4}\cdots$,
with $\Spm=\Sx\pm\iu\Sy$, in particular biaxial systems.


\subsubsection{$\X$ for ladder perturbations
\\
(harmonic oscillators and uniaxial magnets)}

The general expression for the susceptibility obtained by plugging the
$\Z_{i}$ from Eq.~(\ref{Z:pert:expl}) into
$\X = \beta \big[ (\Z_{2}/\Z_{0}) - (\Z_{1}/\Z_{0})^{2} \big]$,
simplifies when the coupled observable $\VQ$ gives, at most,
transitions to adjacent levels: $\m\to\m$ and $\m\to\m\pm1$.
Then one can write
%
\begin{equation}
\label{VQ:ladder}
\VQ
=
\hat{\bd}
\cdot
\J
=
\bd_{0}
\Sco
+
\half
\big(
\bd_{+}
\Scm
+
\bd_{-}
\Scp
\big)
\;,
\end{equation}
with the following action on the unperturbed basis
%
\begin{equation}
\label{ladder:action}
\Sco
\mket
=
\lf_{\m}^{0}
\mket
\qquad
\Spm
\mket
=
\lf_{\m}^{\pm}
|\m\pm1\rangle
\;.
\end{equation}
The harmonic oscillator corresponds to no central term
$\lf_{\m}^{0}=0$ and the creation-destruction factors
$\lf_{\m}^{\pm}=[(\m+\half)\pm\half]^{1/2}$.
For spin problems, the coefficients are $\lf_{\m}^{0}=\m$ and the
custom angular-momentum ladder factors
$\lf_{\m}^{\pm}=\sqrt{\SSp-\m(\m\pm1)}$.~%
\footnote{
Note that here we are already thinking of an unperturbed
$\Hs=\Hs(\Sz)$, otherwise the standard basis, where $\J$ has the
ladder properties~(\ref{ladder:action}), does not diagonalize $\Hs$,
and we would have to resort back to the more
general~(\ref{X:unbiased}).
} 

For these ``ladder perturbations'' the terms with $\bd_{0}$ in
$\Z_{2}/\Z_{0}$ and $(\Z_{2}/\Z_{0})^{2}$ can be combined, while the
ladder action~(\ref{ladder:action}) reduces the double sum
$\sum_{\m\,\n}$ to a single sum. 
The susceptibility then reads
%
\begin{eqnarray}
\label{X:ladder}
\X
=
&
\bd_{0}^{2}
&
\frac{\beta}{\Z_{0}}
\sum_{\m}
\e^{-\bEm}\,
\big[
(\lf_{\m}^{0})^{2}-\langle\lf_{\m}^{0}\rangle^{2}
\big]
\nonumber\\
+
&
\tfrac{1}{4}
\bd_{+}\bd_{-}
&
\frac{\beta}{\Z_{0}}
\sum_{\m}
\big[
(\lf_{\m}^{+})^{2}
\eK_{\m,\m+1}
+
(\lf_{\m}^{-})^{2}
\eK_{\m,\m-1}
\big]
\;
\end{eqnarray}
where we have introduced the shorthand
$\eK_{\m,\n}=\e^{-\bEm}\,K_{\m\,\n}$.
It is not difficult to check the symmetry $\eK_{\n,\m}=\eK_{\m,\n}$, from
the detailed-balance property $K(-X)=\e^{-X}\,K(X)$ mentioned above.
This symmetry, together with $\lf_{\m}^{\pm}=\lf_{\m\pm1}^{\mp}$,
leads to the two sums in the transverse $\bd_{+}\bd_{-}$ part being
equal (the sums, not the summands).
Therefore we can keep one of them, replacing the factor $1/4$ in front
by $1/2$.~%
\footnote{
The equality
$\sum_{\m}
(\lf_{\m}^{+})^{2}
\eK_{\m,\m+1}
=
\sum_{\m}
(\lf_{\m}^{-})^{2}
\eK_{\m,\m-1}$
can also be proved directly from the Kubo correlator, as those terms
correspond to
$\int_{0}^{\beta}\!
\drm\kps\,
\Tr\,\big[
\e^{(\beta-\kps)\Hs(\Sz)}\,
\Spm\,
\e^{+\kps\Hs(\Sz)}\,
\Sm_{\mp}
\big]$.
} 


\subsubsection{
final expression/summary for uniaxial magnets
}

We conclude writing explicitly $\X$ for paramagnets with
$\Hs=\Hs(\Sz)$ probed by a field $\propto-\hat{\bd}\cdot\J$.
The unperturbed basis is then the standard basis $\Sz\mket=\m\mket$.
Besides $\blo\bydefl\bd_{z}$ is the direction cosine of the probing
field {\em parallel\/} to the magnet local axis, while
$\btr^{2}=\bd_{+}\bd_{-}$ corresponds to the {\em transverse\/} one.

The susceptibility can then be decomposed into
%
\begin{equation}
\label{X:lo-tr}
\X
=
\blo^{2}\,
\Xlo
+
\btr^{2}\,
\Xtr
\;,
\end{equation}
with the longitudinal \& transverse components given by
%
\begin{eqnarray}
\label{Xlo}
\Xlo
&=&
\frac{\beta}{\Z_{0}}
\sum_{\m}
\e^{-\bEm}\,
\big(
\m^{2}-\langle\m\rangle^{2}
\big)
\\
\label{Xtr}
\Xtr
&=&
\frac{\beta}{2\Z_{0}}
\sum_{\m}
\e^{-\bEm}\,
\lf_{\m}^{2}\,
K_{\m}
\;.
\end{eqnarray}
Here $\Z_{0}=\sum_{\m}\e^{-\bEm}$ and we have done the mentioned
reduction of the duplicated terms (keeping
$\sum_{\m} (\lf_{\m}^{+})^{2} \eK_{\m,\m+1}$)
and simplified some notations
%
\begin{eqnarray}
\label{ladder:kubo:simpl}
\lf_{\m}^{2}
&\bydefl&
(\lf_{\m}^{+})^{2}
=
\SSp-\m(\m+1)
\\
K_{\m}
&\bydefl&
K_{\m,\m+1}
=
K(\beta\tf_{\m\,\m+1})
\;.
\end{eqnarray}
Thus the Kubo function $K(X)=(\e^{X}-1)/X$ enters evaluated at the
``upward'' transition frequency $\tf_{\m\,\m+1}=\el_{\m}-\el_{\m+1}$.
The rest of the article will consist essentially of examples and
application of Eqs.~(\ref{Xlo}) and~(\ref{Xtr}).


\paragraph*{
some remarks.
}

Before closing this section, two remarks are in order.
First, the longitudinal $\Xlo$ can be obtained directly from
derivatives of the unperturbed partition function $\Z_{0}$, with
respect to the static field, as it is well known.
In our notation, when $\Hs=\Ham_{\rm a}(\Sz)-\Sz\,\Bz$, the levels are
$-\bEm=-\bEm^{\rm a}+\m\,\y$, with the dimensionless variable $\y=\beta\Bz$.
Then
$\Z_{0}=\sum_{\m}\e^{-\bEm^{\rm a}+\m\,\y}$, 
whence
$\Z_{0}\llangle\m^{k}\rrangle=\drm^{k}\Z_{0}/\drm\y^{k}$ 
for any moment $\llangle\m^{k}\rrangle$.
As a result
%
\begin{equation}
\label{}
\Mz
=
\frac{1}{\Z_{0}}
\frac{\drm\Z_{0}}{\drm\y}
\qquad
\MMz
-
\Mz^{2}
=
\frac{\drm\Mz}{\drm\y}
\;,
\end{equation}
from which $\Xlo$ follows as $\Xlo=\beta\,\drm\Mz/\drm\y$ with
$\y=\beta\Bz$.

The last remark is on angular behavior.
The susceptibility proper is a tensor quantity $\X_{ij}$ relating two
vectors (magnetic moment and probing field).
What we have been using throughout is the projected scalar form
$\X
\bydefl
\sum_{ij}\X_{ij}\bd_{i}\bd_{j}
=
\X_{zz}\bd_{z}^{2}+\X_{xx}(\bd_{x}^{2}+\bd_{y}^{2})$,
where the last form holds for uniaxial symmetry.
Then, for a system of non-interacting spins with a distribution of
axes orientations, the scalar $\X$ follows from the corresponding
angular averages $\overline{\blo^{2}}$ and $\overline{\btr^{2}}$.
This can be done, for example, when there is no bias field singling
out a preferred direction; then one has
%
\begin{equation}
\label{X:ran}
\Xran
=
\tfrac{1}{3}
\Xlo
+
\tfrac{2}{3}
\Xtr
\;,
\end{equation}
for the susceptibility of a system with anisotropy axes distributed at
random (powder sample, liquids, etc.).


\section{paramagnets}
\label{sec:PM}

In this section we will check the formulas of Sec.~\ref{derivX}
particularizing them to simple paramagnets.
We will consider isotropic spins $\forall\Sm$ and anisotropic spins
with $\Sm=1$ and $\Sm=3/2$.
The susceptibility for anisotropic problems is typically obtained
using Van Vleck's method, solving the eigenvalue problem in presence
of the probing field \cite{carlin,kahn}.
We will plainly recover those standard formulas of molecular magnetism
textbooks (bypassing diagonalization) and minimally extend them by
including longitudinal bias fields.
Here we will also see the first examples of the $\X(T)$ phenomenology
that will be discussed later on for general~$\Sm$.~%
\footnote{
The reader may well wonder why we have not used Bose/Fermi statistics
for integer/half-integer $\Sm$.
First, the only basic statistics is Gibbs--Boltzmann, with Bose/Fermi
distributions as particular/worked cases of
$\dm\propto\e^{-\beta\Htot}$ for the ideal gas of {\em
indistinguishable\/} bosons or fermions.
But we implicitly bypass indistinguishability by locating/labeling
each spin at a given lattice point (localized moments' magnetism)
\cite[Ch.~3.1]{white}~\cite[Ch.~1.2]{pake}.
This is the same approximation routinely used in solid state physics
and quantum chemistry, where one does not symmetrize/antisymmetrize the
states with respect to the nuclei exchange; 
an approximation grounded on the non-overlapping of the ions'
wave-functions for sufficiently localized states.
} 
%


\subsection{isotropic spin (Brillouin $\X$)}

This is the simplest paramagnetic problem \cite{white,pake}, with
Hamiltonian $\Hs=-\Sz\,\Bz$ and spectrum $-\bEm=\m\,\y$.
The energy levels are equispaced $\beta\tf_{\m\,\m+1}=\y$, yielding a
$\m$-independent Kubo factor $K(\y)$.
Besides, the partition function 
%
\begin{equation}
\label{Z:Siso}
\Z_{0}
=
\sum_{\m=-\Sm}^{\Sm}
\e^{\m\,y}
=
\frac{\sh[(\Sm+\half)\,\y]}{\sh(\half\,\y)}
\;,
\qquad
\y=\Bz/\kT
\;,
\end{equation}
follows readily by summing the geometric series
$\sum_{\m}(\e^{\y})^{\m}$.


\subsubsection{longitudinal susceptibility}

As discussed above, the longitudinal response follows entirely from
$\Bz$-derivatives of $\Z_{0}$.
The first moment reads (magnetization)
%
\begin{equation}
\label{Mz:brillouin}
\Mz
=
(\Sm+\half)\,
\cth[(\Sm+\half)\,\y]
-
\half\,
\cth(\half\,\y)
\;,
\end{equation}
with the right-hand side defining the Brillouin function.
One further derivative, using $(\cth x)'=1-\cth^{2}x$, gives the
susceptibility $\Xlo=\beta\,\drm\Mz/\drm\y$ as
%
\begin{equation}
\label{X:brillouin}
\Xlo
=
\beta\SSp
-
\beta
\left[
(\Sm+\half)^{2}
\cth^{2}[(\Sm+\half)\,\y]
-
\tfrac{1}{4}\,
\cth^{2}(\tfrac{1}{2}\,\y)
\right]
\;.
\end{equation}
In the weak-field limit, one can use $\cth^{2}x\simeq1/x^{2}+2/3$ to
show that $\Xlo$ duly reduces to the {\em Curie law}
%
\begin{equation}
\label{X:curie}
\Xlo
\quad
\stackrel{\scriptstyle \y\to0}
{\longrightarrow}
\quad
\Xc
=
\tfrac{1}{3}\beta\SSp
\;.
\end{equation}
This famous $1/\kT$ dependence was found experimentally by Pierre
Curie in 1895 \cite[Ch.~3]{white} and it has been fruitfully exploited
for calibration and thermometry in the low temperature world.
It expresses the decrease of the response with increasing $T$ due to
the thermal misalignment of the dipole moments away from the probing
field direction.


\subsubsection{transverse susceptibility}

From the general Eq.~(\ref{Xtr}) for $\Xtr$, plus the equispaced
$-\bEm=\m\,\y$ for isotropic spins, one forms
%
\begin{equation}
\Xtr
=
\frac{\beta}{2}
\sum_{\m}
\frac{\e^{\m\,\y}}{\Z_{0}}\,
\lf_{\m}^{2}\,
K(\y)
\;.
\end{equation}
The $\m$-independent Kubo term $K(\y)=(\e^{\y}-1)/\y$ can be taken out
of the sum, which can be done explicitly:
%
\begin{equation}
\sum_{\m}
\frac{\e^{\m\,\y}}{\Z_{0}}
\big[
\SSp-\m(\m+1)
\big]
=
\big[
\cth(\half\,y)
-1
\big]
\,
\Mz
\;.
\end{equation}
But the square bracket can be written as $2/(\e^{\y}-1)$, which
combined with the Kubo factor leaves the simple form
$\Xtr=(\beta/\y)\Mz$.
That is (recall $\y=\beta\Bz$)
%
\begin{equation}
\label{X:tr:iso}
\Xtr
=
\frac{\Mz}{\Bz}
\quad
\stackrel{\scriptstyle \y\to0}
{\longrightarrow}
\quad
\tfrac{1}{3}\beta\SSp
\;,
\end{equation}
which also recovers the correct Curie limit at low fields, as expected
from the restoration of isotropy.%
\footnote{
For $\Sm=1/2$ (the two-level system), one has
$\Mz=\half\,\thrm(\half\,\y)$
(population difference), so that
$\Xlo=\tfrac{1}{4}\beta/\ch\!^{2}(\half\,\y)$,
while
$\Xtr=\thrm(\half\,\y)/2\Bz$.
Both duly give
$\Xc=\tfrac{1}{4}\beta$
at zero field [$\SSp=1/4$].
In the opposite, classical limit $\Sm\gg1$, one has
$\Xlo=\beta\mu^{2}L'(\xi)$ and $\Xtr=(\mu/\Bz)L(\xi)$ for isotropic
superparamagnets, with the Langevin magnetization
$L(\xi)=\cth\xi-1/\xi$ and $\xi=\mu\Bz/\kT$;
see for instance \cite[Eq.~(3.74)]{gar2000}.
} 

It is important to remark that $\Xtr$ does not follow from the
transverse fluctuations of the spin, as one might naively expect.
Indeed, using $\Sx^{2}+\Sy^{2}=\SSp-\Sz^{2}$ one gets for
$\Hs=-\Sz\,\Bz$
%
\begin{equation}
\label{X:Sx:Sy}
\beta\langle\Sx^{2}\rangle
=
\beta\langle\Sy^{2}\rangle
=
\half\,y\,
\cth(\half\,y)
\times
\frac{\Mz}{\Bz}
\neq
\Xtr
\;.
\end{equation}
They only agree at high temperature, $\y=\beta\Bz\ll1$, where the
extra prefactor $\half\,y\,\cth(\half\,y)\to1$.
However, at low temperature, $\Xtr\to S/\Bz$ (following saturation of
the magnetization $\Mz\to\Sm$), whereas $\beta\langle\Sx^{2}\rangle\to
S/2\kT$, which can grow without bound as $T\to0$.
Two different behaviors indeed.%
\footnote{
The unbounded $\beta\langle\Sx^{2}\rangle$ can be seen as due to
``zero point fluctuations'', since $\Sx^{2}+\Sy^{2}\to
\SSp-\Sm^{2}=\Sm$, is different from zero even for ``fully aligned''
$\Sz\to\Sm$; the actual $\Xtr$ corrects for this, and leaves the
response induced by the probing field.
} 


\subsection{anisotropic spins}

When a paramagnetic ion is embedded in a molecule or a solid, the spin
finds preferred orientations which depend on the symmetries of its
neighborhood.
This magnetic anisotropy can be described by adding to the spin
Hamiltonian a term $\Ham_{\rm a}(\J)$ with ``reflection'' symmetry
$\Ham_{\rm a}(-\J)=\Ham_{\rm a}(\J)$ (to respect time-reversal
invariance).
The simplest model is the uniaxial Hamiltonian
%
\begin{equation}
\label{Huni:Bz}
\Hs
=
-\K\,\Sz^{2}-\Sz\,\Bz
\;,
\end{equation}
with $\K$ the anisotropy constant.
To do statistical mechanics we introduce the dimless $\ds=\K/\kT$ \&
$\y=\Bz/\kT$, and write $-\bEm=\ds\,\m^{2}+\y\,\m$.

The minimal mathematical extension from linear to quadratic in $\J$
has important consequences.
First, the spectrum $\el_{\m}=-\K\m^{2}-\Bz\m$ can have a single well
($\K<0$, ``easy-plane'' anisotropy), or it can display a bistable
structure ($\K>0$, ``easy-axis''; see Fig.~\ref{fig:levels}).
Second, the energy levels are no longer equispaced,
$\tf_{\m\,\m+1}=\K(2\m+1)+\Bz$, becoming closer near $\m=0$ (no
``harmonic oscillator'' equispaced simplicity).
In dynamics this gives a multiplicity of precession frequencies
(absorption peaks), tunnel splittings and relaxation rates
\cite{pake,garzue2006,shi80I,zuegar2006}.
But already in the statics we will find a $\m$-dependent Kubo factor
$K(\beta\tf_{\m\,\m+1})$, expressing that
$\e^{-\kps\Sz^{2}}\,\Spm\e^{+\kps\Sz^{2}}$ cannot be written in terms
of $\Spm$ only (another reason underlying the simplicity of isotropic
spins).
Finally, the partition function
$\Z_{0}=\sum_{\m}\e^{\ds\,\m^{2}+\y\,\m}$, with $\ds=\K/\kT$ and
$\y=\Bz/\kT$, cannot be summed explicitly (Gauss type sums), though
simple formulas can be produced for small $\Sm$.~%
\footnote{
Slow dynamics at low $T$ is an important consequence of the
anisotropy.
From the point of view of the many relaxation rates
\cite{shi80I,zuegar2006}, one of the rates is well separated from the
others (the analogue of the lowest non-vanishing eigenvalue in the
framework of the Fokker--Planck equation).
The faster rates correspond to intrawell modes, while the slow mode is
associated with the overbarrier dynamics of the spin; due to thermal
activation it is suppressed exponentially at low $T$, by
$\exp(-\dU/\kT)$.
The phenomenological ``blocking temperature'' is that where the
observational time window $t_{\rm m}$ matches this slow dynamics;
below it, the given technique does not ``record'' anymore equilibrium
properties.
} 
%


\subsubsection{susceptibilities for $\Sm=1$}

Spin one is the first case giving new phenomenology;
in the case $\Sm=1/2$ the term $-\K\,\Sz^{2}$ produces a uniform level
shift, so that $\X$ is the same as that of the isotropic spin
discussed above.

For $\Sm=1$ at zero field, there are two facing energy levels
$\m=\pm1$ (minima for $\K>0$) with the level $\m=0$ in between (a
maximum or minimum, depending on the sign of $\K$; top left panel in
Fig.~\ref{fig:levels}).
In Table~\ref{table:S1} we have collected the energy levels, ladder
factors, and transition frequencies required to calculate the
equilibrium response.
In terms of $\ds=\K/\kT$ and $\y=\Bz/\kT$, the partition
function~(\ref{dm:Z}) reads
%
\begin{equation}
\label{Z:S1}
\Z_{0}
=
1
+
2\,
\e^{\ds}\,
\ch\y
\;,
\qquad
(\Sm=1)
\;,
\end{equation}
which enters in both $\Xlo$ and $\Xtr$.


\paragraph{
longitudinal susceptibility $\Xlo(\Sm=1)$.
}
Differentiating $\Z_{0}$ one gets the magnetization
$\Mz=\Z_{0}^{-1}\drm\Z_{0}/\drm\y$, which can be cast in the
suggestive form
%
\begin{equation}
\label{mz:S1}
\Mz
=
\frac
{\sh\y}
{\ch\y+\half\,\e^{-\ds}}
\;.
\end{equation}
The $\m=0$ level does not contribute to the numerator
$\sum\m\,\e^{-\bEm}$, but occupies a ``phase space'' 
$\half\,\e^{-\ds}$.
When for $\ds\gg1$ the level $\m=0$ gets thermally depopulated, one
recovers $\Mz\simeq\thrm\y$, as in a two-level system.
\begin{table}[!tb]
\begin{center}
\begin{tabular}{r|ccc}
$\m$
&
$-\bEm$
&
$\lf_{\m}^{2}$
&
$\beta\tf_{\m\,\m+1}$
\\
\hline
$-1$
&
$\ds-\y$
&
$2$
&
$-\ds+\y$
\\
$0$
&
$0$
&
$2$
&
$+\ds+\y$
\\
$+1$
&
$\ds+\y$
&
$0$
&
---
\\
\end{tabular}
\end{center}
\caption{
Energy levels
$-\bEm=\ds\,\m^{2}+\y\,\m$,
ladder factors $\lf_{\m}^{2}=2-\m(\m+1)$,
and transition frequencies
$\beta\tf_{\m\,\m+1}=\beta(\el_{\m}-\el_{\m+1})=\ds(2\m+1)+\y$,
for $\Sm=1$.
Note $\SSp=2$ and $\tfrac{1}{3}\SSp=2/3$.
}
\label{table:S1}
\end{table}

Differentiating $\Mz$ gives the fluctuations
$\MMz-\Mz^{2}=\drm\Mz/\drm\y$, which times $\beta=1/\kT$ gives the
longitudinal susceptibility
%
\begin{equation}
\label{X:lo:S1}
\Xlo
=
\beta
\,
\frac
{1+\half\,\e^{-\ds}\,\ch\y}
{(\ch\y+\half\,\e^{\ds})^{2}}
\;,
\qquad
(\Sm=1)
\;.
\end{equation}
Again, $\ds\gg1$ leads to the two-level type susceptibility
$\Xlo\simeq\beta/\ch\!^{2}\y$.


\paragraph{
transverse susceptibility $\Xtr(\Sm=1)$.
}
Picking from Table~\ref{table:S1}, plugging in Eq.~(\ref{Xtr}), and
playing with $K(-X)=\e^{-X}\,K(X)$, we can write the transverse
response as
$\Xtr=(\beta/\Z_{0})\big[K(\ds-\y)+K(\ds+\y)\big]$.
Alternatively, we can unfold the Kubo functions and use
$(\ds\pm\y)/\beta=\K\pm\Bz$ obtaining
%
\begin{equation}
\label{X:tr:S1}
\Xtr
=
\frac{1}{\Z_{0}}
\bigg[
\frac{\e^{\ds-\y}-1}{\K-\Bz}
+
\frac{\e^{\ds+\y}-1}{\K+\Bz}
\bigg]
\;,
\qquad
(\Sm=1)
\;.
\end{equation}
As mentioned before, $\Xtr$ remains finite when adjacent levels become
degenerate (e.g., when $\Bz=\pm\K$).
%
%
Note finally that when $\ds\to0$, we have
%
\begin{equation}
\label{X:tr:iso:S1}
\Xtr
\simeq
\frac{1}{\Bz}
\frac{\sh\y}{\ch\y+\half}
=
\frac{\Mz|_{\Sm=1}}{\Bz}
\;,
\end{equation}
duly recovering the transverse susceptibility~(\ref{X:tr:iso}) for
isotropic $\Sm=1$ spins.
\begin{figure}[!t]
\centerline{
\includegraphics[width=7.cm]{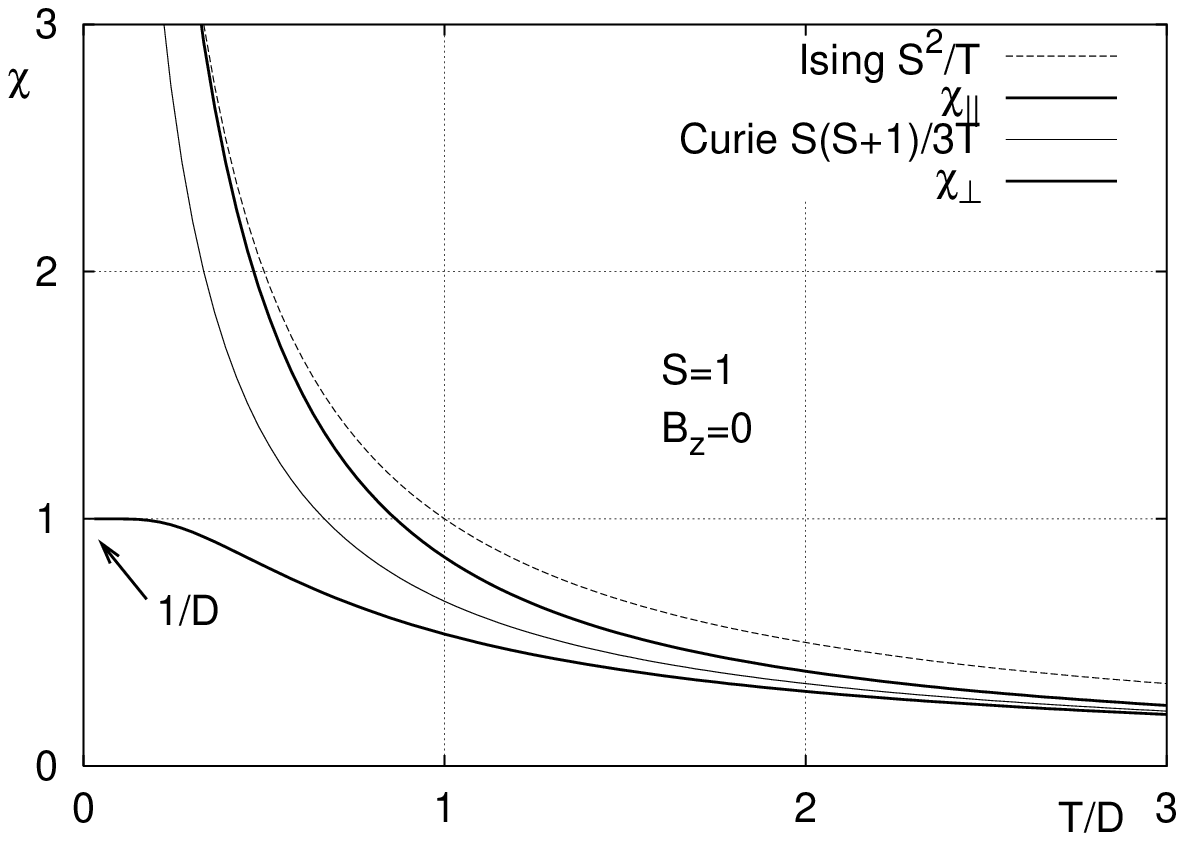}
\hspace*{-2.ex}
\includegraphics[width=7.cm]{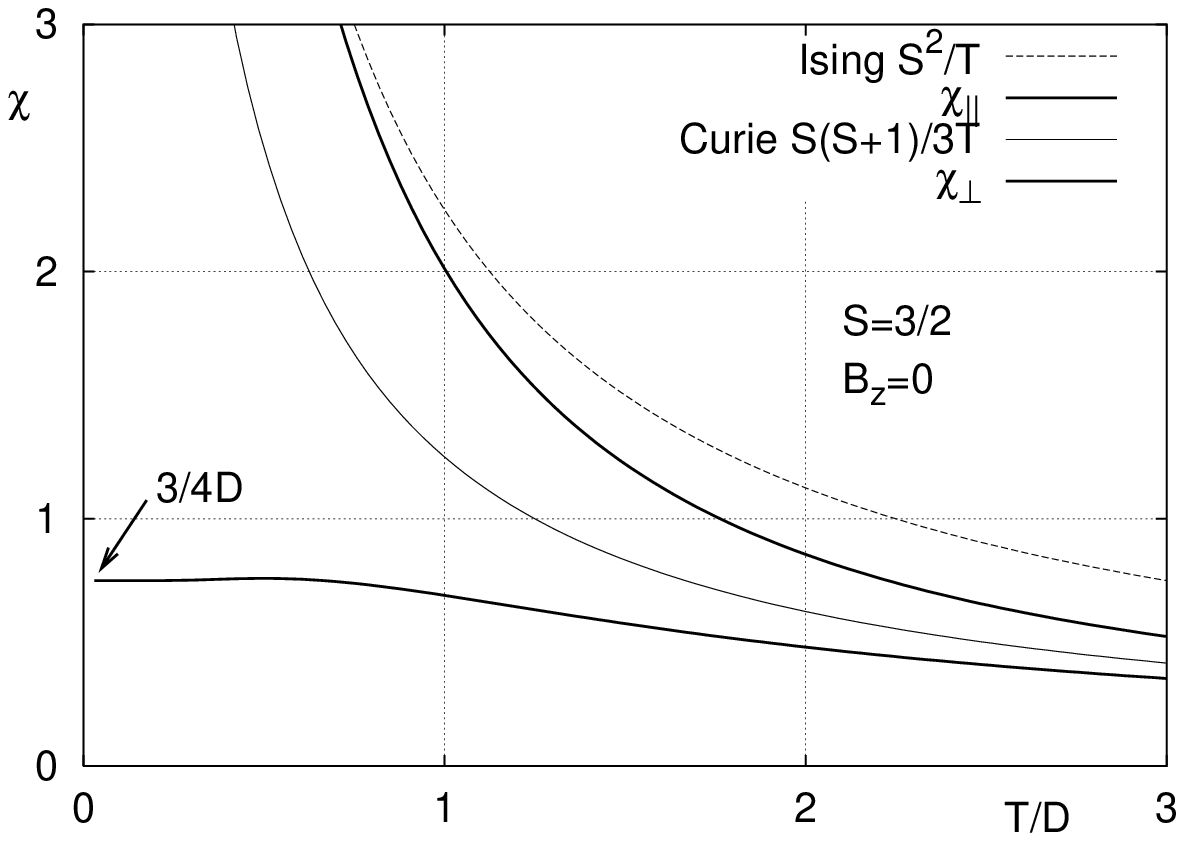}
}
\centerline{
\includegraphics[width=7.cm]{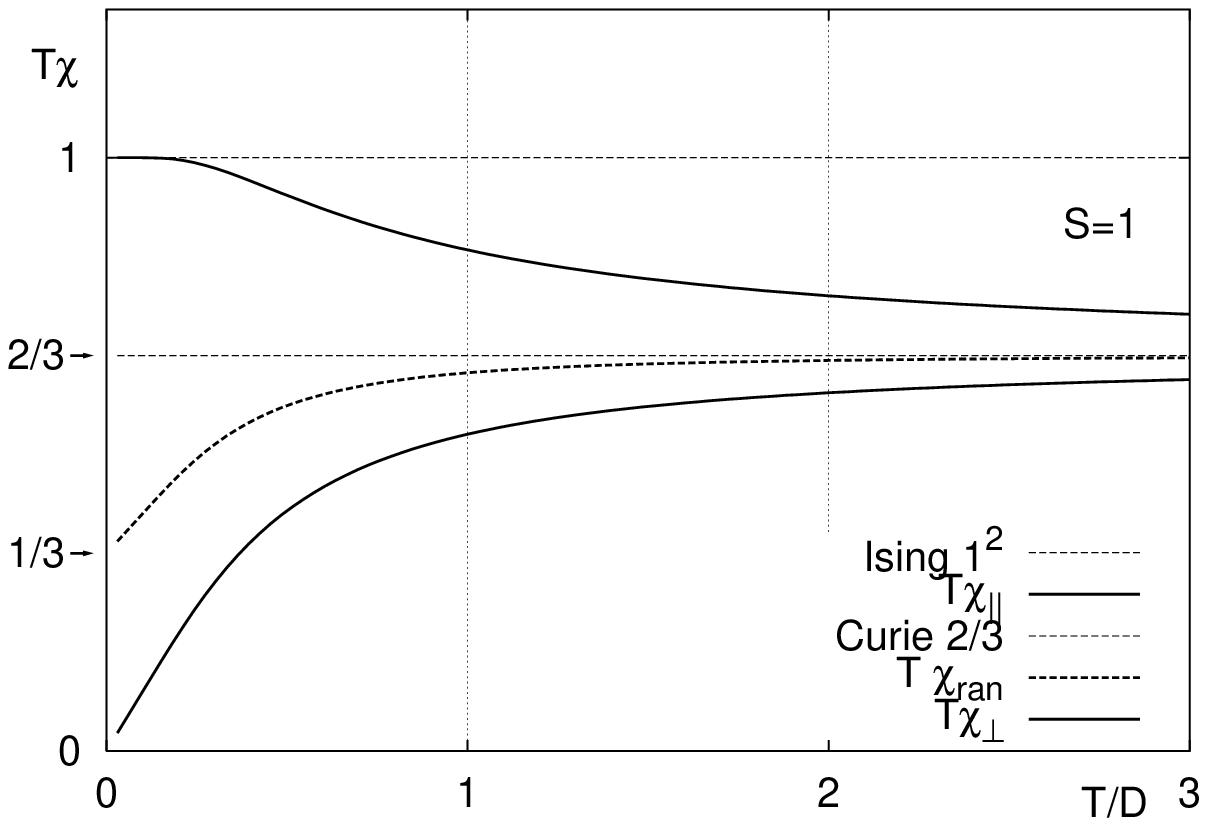}
\hspace*{-2.ex}
\includegraphics[width=7.cm]{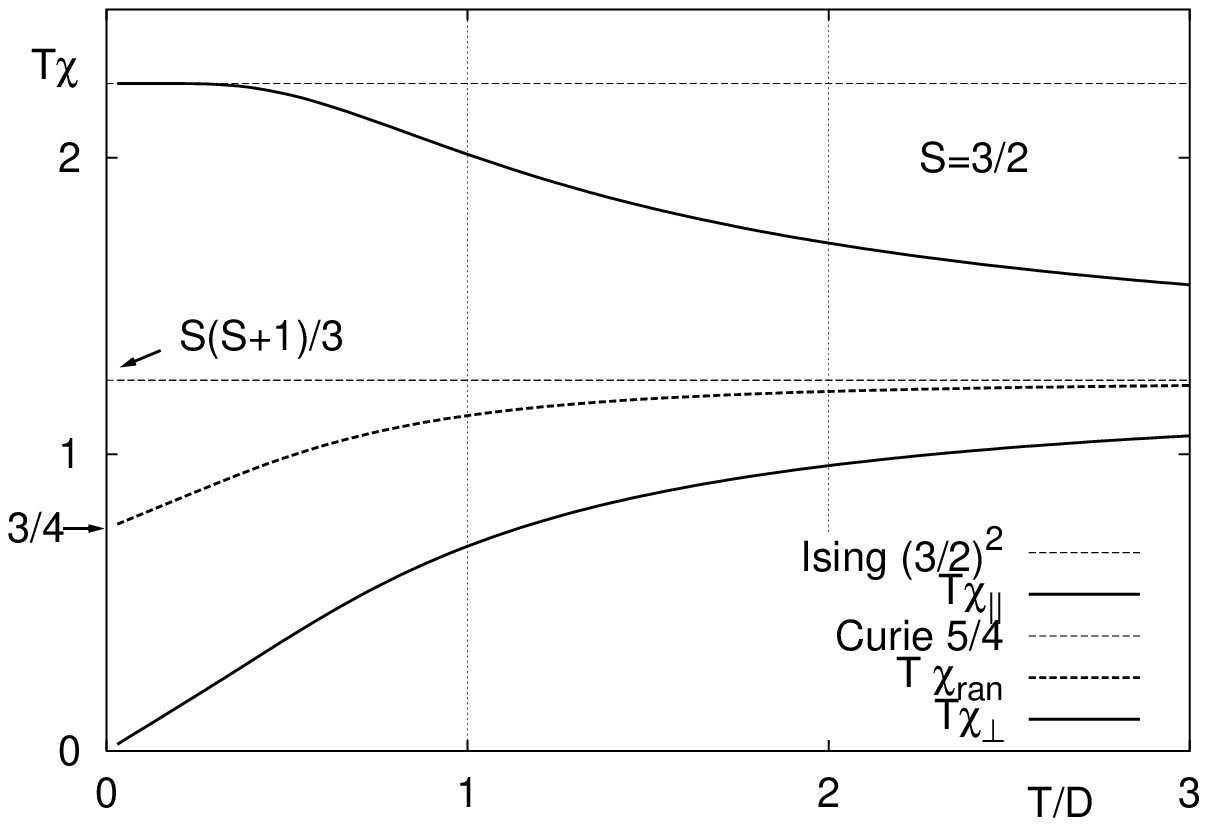}
}
\caption{
Susceptibilities vs.\ temperature of spins $\Sm=1$ (left panels) and
$\Sm=3/2$ (right) with easy-axis anisotropy $\K=1$ (see the spectra in
Fig.~\ref{fig:levels}).
Top panels:
Longitudinal and transverse susceptibility vs.\ $T$.
The Curie law $\SSp/3\kT$ (thin solid lines) is approached by both $\Xlo$
 and $\Xtr$ at high $T$.
At low temperature $\Xlo$ goes over the 2-state behavior $\Sm^{2}/\kT$
(dashed thin lines).
The transverse susceptibility tends to the constant
$\Xtr=1/[2\K(1-1/2\Sm)]$ at low $T$ [Eq.~(\ref{Xtorque})].
Bottom panel:
Susceptibilities plotted as $\kT \X$, including the average for axes
distributed at random, $\Xran=\frac{1}{3}\Xlo+\frac{2}{3}\Xtr$, which shows
deviation from isotropy at lower temperatures.
}
\label{fig:X:lo-tr-ran}
\end{figure}


\paragraph[$T$ dependence of $\Xlo$ and $\Xtr$ for $\Sm=1$]
{
temperature dependence of $\Xlo$ and $\Xtr$ for $\Sm=1$.
}
We will set $\Bz=0$ to describe briefly the features of the
$T$-dependent susceptibility.
The above expressions, Eqs.~(\ref{X:lo:S1}) and~(\ref{X:tr:S1}), then
reduce to (cf.\ Ref.~\cite[Eq.~(2.9)]{carlin})
%
\begin{equation}
\label{X:S1}
\Xlo
=
\frac{2}{\kT}
\,
\frac{1}{2+\e^{-\K/\kT}}
\;,
\qquad
\Xtr
=
\frac{2}{\K}
\,
\frac{1-\e^{-\K/\kT}}{2+\e^{-\K/\kT}}
\;,
\qquad
(\Sm=1)
\;.
\end{equation}
These susceptibilities are plotted in Fig.~\ref{fig:X:lo-tr-ran}.
At high temperature both go over the Curie curve $\Xc=\SSp/3\kT$.
This can be seen from the formulas too, as $|\K/\kT|\ll1$ yields
$\Xlo\simeq2/3\kT$ and $\Xtr\simeq2/3\kT$ (now $\SSp=2$).
As the temperature decreases, $\Xtr$ shows no peaks and tends to the
constant response $\Xtr=1/\K$ (the analogue to the classical torque
susceptibility $\Xtr\sim\half\partial_{\theta}^{2}E$
\cite{chikazumi}).
The longitudinal response, on the other hand, goes over the 2-state
asymptote $\Xlo\simeq1/\kT=\Sm^{2}/\kT$ at low temperatures.
Therefore, as $\Sm^{2}>\SSp/3$, the increase of $\Xlo$ is faster than
$1/T$ in the intermediate range.

We mentioned the routine use of $\X(T)$ curves as characterization
tool \cite{maurer}.
For instance deviations from $1/T$ laws can indicate spin-spin
interactions (think of the mean-field $\X\sim1/(T-\theta)$ when
observed over a short $T$ interval).
However, one should be careful with not overlooking other possible
sources of deviation, as those due to the anisotropy just discussed.


\subsubsection{susceptibilities for $\Sm=3/2$}

We move on to the next spin value.
The anisotropy term $-\K\,\Sz^{2}$ gives two pairs of degenerate
levels, without a central maximum or minimum (see
Fig.~\ref{fig:levels}, top right).
The pair $|\m|=3/2$ is above or below the pair $|\m|=1/2$ depending on
the sign of the anisotropy constant $\K$.

We have put up in Table~\ref{table:S1p5} the relevant quantities to
calculate the susceptibilities of this 4-level system.
From there we first compose the partition function ($\ds=\beta\K$,
$\y=\beta\Bz$)
%
\begin{equation}
\label{Z:S1p5}
\Z_{0}
=
2
\big[
\ch(\y/2)
+
\e^{2\ds}\,
\ch(3\y/2)
\big]
\;,
\qquad
(\Sm=3/2)
\;,
\end{equation}
which will be needed in both the longitudinal and in the transverse
response.


\paragraph{
longitudinal $\Xlo(\Sm=3/2)$.
}
The $\y$-derivative of the above $\Z_{0}$ produces the magnetization
%
\begin{equation}
\label{mz:S1p5}
\Mz
=
\frac{1}{2}
\frac
{\sh(\y/2)+3\,\e^{2\ds}\,\sh(3\y/2)}
{\ch(\y/2)+\e^{2\ds}\,\ch(3\y/2)}
\;.
\end{equation}
As a check, one finds at low $T$ two different 2-level type responses
depending on the sign of $\ds$, namely, $\Mz\simeq(3/2)\thrm(3\y/2)$
for positive $\K$ and $\Mz\simeq(1/2)\thrm(\y/2)$ for negative, with
the right spin values, $3/2$ and $1/2$ respectively.
\begin{table}[!tb]
\begin{center}
\begin{tabular}{r|rcr}
$\m$
&
$-\beta\bar{\el}_{\m}$
&
$\lf_{\m}^{2}$
&
$\beta\tf_{\m\,\m+1}$
\\
\hline
$-3/2$
&
$2\ds-3\y/2$
&
$3$
&
$-2\ds+\y$
\\
$-1/2$
&
$-\y/2$
&
$4$
&
$+\y$
\\
$+1/2$
&
$+\y/2$
&
$3$
&
$+2\ds+\y$
\\
$+3/2$
&
$2\ds+3\y/2$
&
$0$
&
---
\\
\end{tabular}
\end{center}
\caption{
$\Sm=3/2$ energy levels
$-\bEm=\ds\,\m^{2}+\y\,\m$,
ladder factors $\lf_{\m}^{2}=15/4-\m(\m+1)$
and transition frequencies
$\beta\tf_{\m\,\m+1}=\beta(\el_{\m}-\el_{\m+1})=\ds(2\m+1)+\y$.
Here $\SSp=15/4$ and $\tfrac{1}{3}\SSp=5/4$, and to spare
exponentials we have shifted all energies by $\ds(1/2)^{2}$, that is
$\beta\bar{\el}_{\m}\bydefl\bEm+\ds/4$.
}
\label{table:S1p5}
\end{table}

One more derivative gives the longitudinal fluctuations entering in
the susceptibility $\Xlo=\beta\MMz-\Mz^{2}$:
%
\begin{equation}
\label{X:lo:S1p5}
\Xlo
=
\frac{\beta}{4}
\,
\frac
{1+2\,\e^{2\ds}(4\ch\y+\ch 2\y)+9\,\e^{4\ds}}
{[\ch(\y/2)+\e^{2\ds}\,\ch(3\y/2)]^{2}}
\;,
\qquad
(\Sm=3/2)
\;.
\end{equation}
Consistently with the above magnetizations, $\ds\gg1$ gives
$\Xlo\simeq\beta(3/2)^{2}/\ch\!^{2}(3\y/2)$
while $\ds\ll-1$ yields
$\Xlo\simeq\beta(1/2)^{2}/\ch\!^{2}(\y/2)$
as low-temperature 2-level asymptotics.


\paragraph{
transverse $\Xtr(\Sm=3/2)$.
}
We next assemble the transverse susceptibility from the quantities of
Table~\ref{table:S1p5}
%
\begin{equation}
\label{X:tr:S1p5}
\Xtr
=
\frac{\beta}{2\Z_{0}}
\big[
3\,\e^{-\y/2}\,
K(2\ds-\y)
+
4\,\e^{-\y/2}\,
K(+\y)
+
3\,\e^{+\y/2}\,
K(2\ds+\y)
\big]
\;,
\quad
(\Sm=3/2)
\;.
\end{equation}
with the custom $K(X)=(\e^{X}-1)/X$.
We have played again with the detailed-balance property
$K(-X)=\e^{-X}\,K(X)$ to arrive at a compact form
[e.g., note that the middle term is either
$\e^{-\y/2}\,K(+\y)=\e^{+\y/2}\,K(-\y)=\sh(\y/2)/(\y/2)$].


\paragraph[$T$ dependence of $\Xlo$ and $\Xtr$ for $\Sm=3/2$]
{
temperature dependence of $\Xlo$ and $\Xtr$ for $\Sm=3/2$.
}
In the unbiased $\Bz=0$ case, Eqs.~(\ref{X:lo:S1p5})
and~(\ref{X:tr:S1p5}) reduce to
%
\begin{equation}
\label{X:lo:tr:unb:S1p5}
\Xlo
=
\frac{\beta}{4}
\,
\frac
{\e^{-2\ds}+10+9\,\e^{+2\ds}}
{4\ch^{2}\ds}
\;,
\qquad
\Xtr
=
\frac{\beta}{2(1+\e^{2\ds})}
\big[
3\,
K(2\ds)
+
2
\big]
\;.
\end{equation}
Both approach the Curie curve $\Xc=5/4\kT$ when $\ds=\K/\kT\to0$, as
seen in the high-temperature range of Fig.~\ref{fig:X:lo-tr-ran} (now
$\SSp=15/4$).
Again, at low temperature $\Xlo$ has to catch up with the 2-level
asymptote $\Xlo\simeq\Sm^{2}/\kT$ (with $\Sm=3/2$ for $\K>0$),
increasing faster than $1/T$ in the crossover range.

The transverse $\Xtr$ tends to a constant value as $T\to0$, but it
displays a small and broad maximum around $\kT\sim\K/2$, which was
absent in $\Sm=1$ (and hardly visible here).
The maximum can be obtained from the alternative form
$\Xtr
=
\frac{3}{4\K}
\,
[\thrm\ds
+
\tfrac{2}{3}\,
\ds\,
(1-\thrm\ds)]$,
where $\ds\,(1-\thrm\ds)$ adds a small bump to the monotonous
$\thrm\ds$ as $T$ decreases ($\ds=\K/\kT$).
%

We close with the susceptibility for the ensemble with anisotropy axes
oriented at random.
Figure~\ref{fig:X:lo-tr-ran} shows that
$\Xran=\frac{1}{3}\Xlo+\frac{2}{3}\Xtr$
matches the isotropic Curie curve over a wider $T$ range than $\Xlo$
and $\Xtr$;
actually down to $\kT\sim\K$--$2\K$, where it eventually deviates
downwards.
This behavior was also shown by $\Sm=1$, and will be addressed in the
next section for larger $\Sm$ values.


\section{superparamagnets}
\label{sec:SPM}

Discussing the $T$ dependence of the susceptibility of anisotropic
spins $\Sm=1$ and $\Sm=3/2$ we have come across some features of the
curves that we would like to study more systematically, for several
$\Sm$.
We want to check if those behaviors are specific of some spin values,
how they evolve with $\Sm$, and whether they survive the
classical/continuum limit $\Sm\to\infty$.

We have in mind three features: (1) the crossover of $\Xlo$ from the
Curie law $\Xc=\third\beta\SSp$ toward the 2-state response as $T$ is
lowered, (2) the peak in the transverse $\Xtr(T)$ around the
anisotropy constant $\K$, and (3) the deviation of the orientationally
randomized $\Xran$ from Curie at low temperature.%
\footnote{
One may think that (1) and (3) are the same, but in classical
superparamagnets there is crossover in $\Xlo$ but no deviation of
$\Xran$ from Curie \cite{garjonsve2000}.
} 

Technically, the study as a function of $\Sm$ is eased by the compact
expressions of Sec.~\ref{derivX}.
We just need an algorithm building a table like those of
Sec.~\ref{sec:PM}, for a given $\Sm$, and feeding the formulas for
$\Xlo$ and $\Xtr$ [Eqs.~(\ref{Xlo}) and~(\ref{Xtr})] with the entries
$\el_{\m}$, $\lf_{\m}^{2}=\SSp-\m(\m+1)$, and
$\tf_{\m\,\m+1}=\el_{\m}-\el_{\m+1}$.


\subsection{
scaling with $\Sm$
}
\label{scaling:S}

To compare results for different $\Sm$ meaningfully, we must specify
which parameters are kept constant, or how they are scaled with $\Sm$.
The same applies to the classical limit.
This kind of specifications is needed in any sensible limit-taking
procedure in physics (thermodynamical limits, continuum limit from
mechanics, or from lattice discretizations, etc.).
Different specifications/scalings give different results, possibly
answering different questions.

We will use the following prescription.
We compare spins with different $\Sm$ but having the same (maximum)
energy.
For the spin Hamiltonian $\Hs=-\K\,\Sz^{2}-\B\,\cdot\J$, this entails
to keep fixed
%
\begin{equation}
\label{scalings}
{\rm scaling:}
\qquad
\K\,\Sm^{2}
=
{\rm const.}
\qquad
B\,\Sm
=
{\rm const.}
\;,
\end{equation}
as the spin is varied.
Then the energy differences between adjacent levels $\tf\sim\K\,\Sm+\Bz$
will decrease as $\tf\sim1/\Sm$, and the levels will approach each other
accordingly.
In Fig.~\ref{fig:levels} we have shown three spins with the same
anisotropy barrier $\dU=\K\Sm^{2}$ but different number of levels ($3$,
$5$ and $9$).
Therefore, with this convention the question we will be answering is
how the number of levels, as a measure of discreteness/quantumness,
would affect the properties observed.

In the rest of this section we will address the three points mentioned
for the uniaxial model $\Hs=-\K\,\Sz^{2}-\Bz\,\Sz$.
Having in mind actual superparamagnets, we use $\K>0$ and set $\Bz=0$
for simplicity.
The quantity $\K\,\Sm^{2}$, being fixed, provides a natural scale of
temperature $\kT/\K\,\Sm^{2}$, with
$\K\,\Sm^{2}\sim10^{2}$--$10^{3}$\,K in many superparamagnets
\cite{panpol93,blupra2004}.
Similarly, the output susceptibility curves will be normalized as
$\X/\SSp$, to facilitate comparisons.
For example, all curves $\kT \X$ would go over $1/3$ at high $T$, and
we can discuss how the low $T$ behavior is modified by the
discreteness of the spectrum.
\begin{figure}[!tbh]
\centerline{
\includegraphics[width=10.cm]{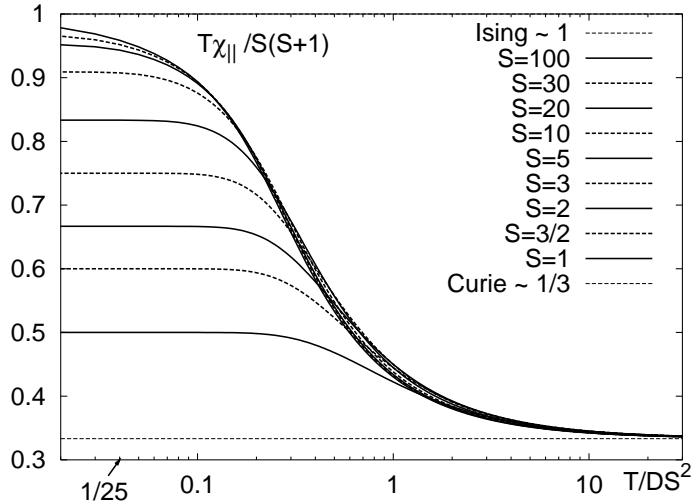} 
}
\caption{
Longitudinal susceptibility $\Xlo$ vs.\ $T$ for several $\Sm$.
Curves presented as $\kT \X/\SSp$, to show the crossover from the
high-$T$ Curie regime, $\kT \X/\SSp=1/3$ (thin dashed line), to the
2-level regime at low temperatures.
The arrow at $\sim1/25$ indicates the lower limit of the observable
equilibrium temperature window due to the onset of non-equilibrium
effects (superparamagnetic blocking; cf.~\cite{ponomarenko}).
}
\label{fig:X:manyS:lo}
\end{figure}


\subsection{
longitudinal response for various $\Sm$
}
\label{Xlo:S}

As already mentioned, understanding well the $T$ dependence of the
susceptibility of the paramagnetic ``ideal gas'' is important because
$\X(T)$ curves are routinely used as indicators of spin-spin
interactions, anisotropies, etc \cite{carlin,kahn,maurer}.

In Sec.~\ref{sec:PM} we saw that for anisotropic $\Sm=1$ and
$\Sm=3/2$, at high enough temperature, the susceptibility approaches
the Curie asymptote $\Xc=\third\beta\SSp$, whereas in the low $T$
range the highly anisotropic 2-state regime $\Xising=\beta\Sm^{2}$
emerges (if you prefer, Heisenberg to Ising crossover).
This is expected $\forall\,\Sm$, because at high enough $T$ the
anisotropy term plays a minor role $\e^{-\bEm}\simeq1$, and the spin
effectively becomes a free (quantum) rotor.
Indeed, arithmetic sums like
$\sum_{k=1}^{N}k^{2}=\third N(N+1)(N+\half)$ 
give
%
\begin{equation}
\label{}
\sum_{\m=-\Sm}^{\Sm}
\m^{2}
\e^{-\bEm}
\simeq
\sum_{\m=-\Sm}^{\Sm}
\m^{2}
=
\third \SSp(2\Sm+1)
\;,
\end{equation}
which divided by $\Z_{0}\simeq\sum_{\m}1=2\Sm+1$ (number of states) produces
the Curie law $\Xc=\third\beta\SSp$, for all $\Sm$.
Then, as $-\bEm=\beta\K\,\m^{2}$, the onset of deviation from isotropy
can be defined as the temperature where $\e^{-\beta\el_{0}}=1$ and
$\e^{-\beta\el_{\Sm}}$ (population of the ``poles'' and the
``equator'') start to differ appreciably.
Say by $5$\,\%, then $\e^{\K\,\Sm^{2}/\kT}\simeq1+\K\,\Sm^{2}/\kT\sim1.05$
gives $\kT/\K\,\Sm^{2}\sim20$.
This estimate holds $\forall\,\Sm$, as seen in the high-$T$ range of
Fig.~\ref{fig:X:manyS:lo}.

In the opposite low $T$ regime, we can approximate $\Xlo$ by using the
lowest levels $\m=\pm\Sm$ (effective 2-state; classically only the
poles populated)
%
\begin{equation}
\label{}
\MMz
=
\frac
{\sum_{\m}\e^{-\bEm}\,\m^{2}}
{\sum_{\m}\e^{-\bEm}}
\simeq
\frac
{\e^{-\beta\el_{-\Sm}}\,(-\Sm)^{2}+\e^{-\beta\el_{\Sm}}\,\Sm^{2}}
{\e^{-\beta\el_{-\Sm}}+\e^{-\beta\el_{\Sm}}}
=
\Sm^{2}
\quad
\leadsto
\quad
\Xlo
\simeq
\beta\Sm^{2}
\;.
\end{equation}
In this low $T$ range, $\kT\Xlo$ shows a plateau
(Fig.~\ref{fig:X:manyS:lo}) which does not bend until the next levels
$\el_{\pm(\Sm-1)}$ become appreciably populated.
This suggests that the leaving this ``Ising plateau'' is governed by
$T_{\rm I}\bydefl\tf_{\Sm,\Sm-1}=\K(2\Sm-1)$
(the energy difference to the first excited pair).
As this energy enters exponentially, the curves' bend would be
negligible at a $T\sim$ five times lower [$\exp(-x)\simeq0$ at
$x\sim4$--$6$].
Indeed in our scaled units
$T_{\rm I}/5\K\,\Sm^{2}=(2\Sm-1)/5\Sm^{2}\sim2/5\Sm$.
This gives a longer plateau the smaller $\Sm$ is, in agreement with
the curves of Fig.~\ref{fig:X:manyS:lo}.
Note that for large $\Sm$, not only is the Ising regime shorter, but
it is reached more slowly when decreasing $T$ (non exponentially).%
\footnote{ 
For classical superparamagnets $\Xlo=\beta\mu^{2}R'/R$ where
$R(\alpha)=\int_{0}^{1}\!\drm z\,\exp(\alpha z^{2})$
is the ``partition function'' with $\alpha=\dU/T$, and $R'=\drm
R/\drm\alpha$
\cite{gar2000,ponomarenko,wes61,raishl75e,shc78,chaayopop85}.
%
%
At low $T$ (large $\alpha$) one can approximate $R'/R\simeq1-1/\alpha$
showing the non-exponential, power-law approach to
$\Xlo\simeq\beta\mu^{2}$.
At high $T$, on the other hand, letting $\alpha\to0$ one has
$R'/R=\int_{0}^{1}\!\drm z\,z^{2}=1/3$,
recovering the classical Curie susceptibility
$\Xlo\simeq\third\beta\mu^{2}$.
} 

With the two estimates above we have characterized the extent of the
temperature crossover for all $\Sm$.
As for the magnitude of the ``excursion'', it just follows from $\kT\Xlo$
evolving from $\third\SSp$ up to $\Sm^{2}$
%
\begin{equation}
\label{excursion}
\frac{\Sm^{2}}{\third\SSp}
=
\frac{3\,\Sm}{\Sm+1}
=
\left\{
\begin{array}{lcl}
3
&&
S\to\infty
\\[0.ex]
3/2
&&
\Sm=1
\end{array}
\right.
\;.
\end{equation}
This is maximum classically, $3$, and decreases with $\Sm$ to get halved
for $\Sm=1$.
This merely reflects the familiar quantum-mechanical fact of
$\vec{\Sm}^{2}$ not having length $\Sm^{2}$, but $\SSp$.

Having characterized the crossover, one could assess the temperature
ranges where approximate modelizations can be employed.
For example, for classical nanoparticles the use of the Curie
susceptibility is widespread (with the corresponding Langevin
magnetization), unfortunately well down to the superparamagnetic blocking
$\kT/\K\,\Sm^{2}\sim0.04$--$0.1$ (arrow in Fig.~\ref{fig:X:manyS:lo}).
Here we see once more that if a rough model is to be chosen, the
2-state model is preferable [with the associated
$\Mz\propto\thrm(\mu\,B/\kT)$], as the pioneers of superparamagnetism
properly did \cite{nee49}.
\footnote{
The folk view associates superparamagnetism with energy barriers
$\dU=\K\,\Sm^{2}$ larger than $\kT$.
But a crude estimate of the over-barrier relaxation time
$\tau=\tau_{0}\exp(\beta\dU)$ gives a pre-factor
$\tau_{0}\sim10^{-7}$--$10^{-8}$\,s for molecular clusters and
$\tau_{0}\sim10^{-10}$--$10^{-12}$\,s for nanoparticles
\cite{blupra2004,panpol93}.
Superparamagnetism is to be observed for measurement times $t_{\rm
m}\gg\tau$.
Then one can have an equilibrium/superparamagnetic temperature range
as wide as $25>\beta\dU\geq 0$, for static measurements $t_{\rm
m}\sim1$--$100$\,s, showing that the folk ascription $1>\beta\dU\geq0$
is hopelessly restrictive.
For example, almost the whole crossover from $\beta\dU\ll1$ to
$\beta\dU\gg1$ can fit in the observational equilibrium temperature
window (we marked its lower limit $\sim1/25$ in the $T$ axis in
Fig.~\ref{fig:X:manyS:lo}).
} 


\subsection{
transverse susceptibility for various $\Sm$
}
\label{Xtr:S}

The transverse response is a process quite different from the
longitudinal one.
$\Xlo$ is about applying a small probe that shifts up and down the
energy levels, and the associated re-population involving
$\exp(-\beta\dU)$ factors.
However, a transverse probe mixes and splits the degenerate levels,
and our intuition of the response gets somewhat lost.

A classical cartoon can be of some assistance.
In it the effect of $\btr$ is trying to rotate $\J$ out of the stable
anisotropy minima (the ``up'' and ``down'' poles) toward some point
in the equator.
The result of this torque can be expressed as
$\Xtr\sim\half\partial_{\theta}^{2}E$.
Thus, the transverse susceptibility is more sensitive to other
features, like energy-well curvatures, rather than to barrier heights,
which renders it as a valuable tool too.
\begin{figure}[!tbh]
\centerline{
\includegraphics[width=10.cm]{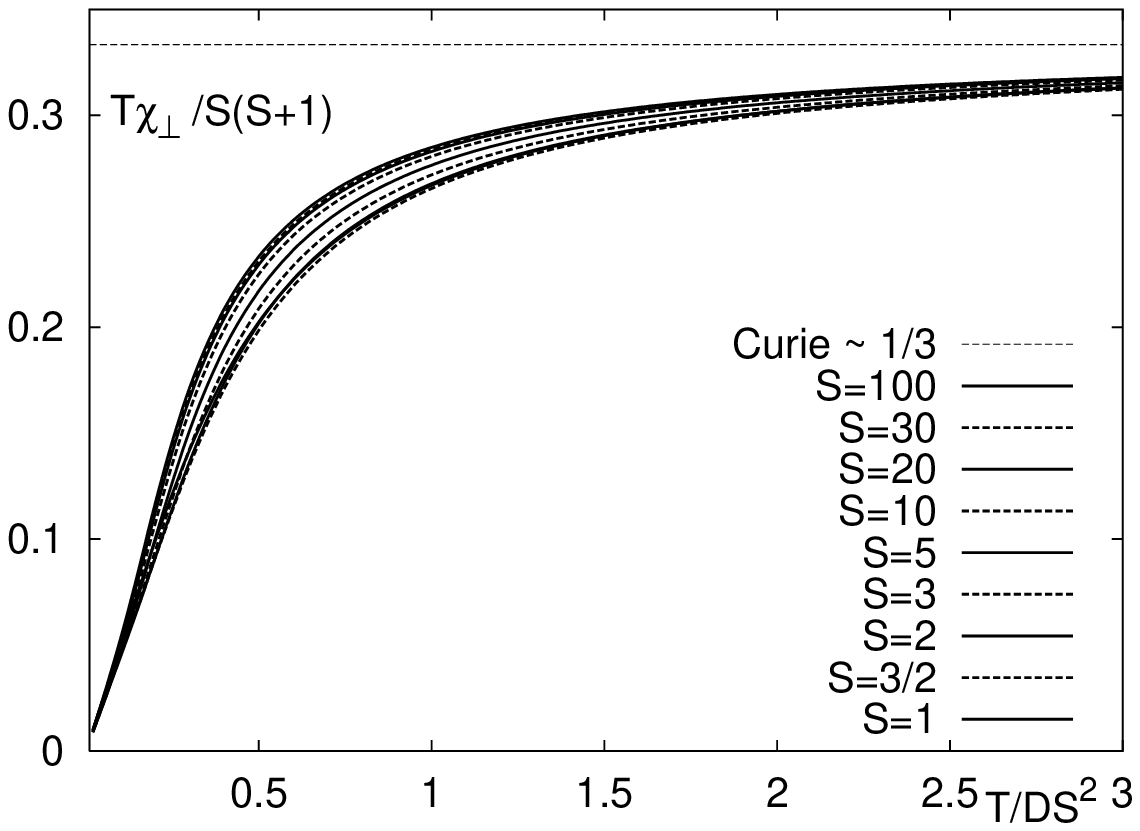}
}
\centerline{
\includegraphics[width=10.cm]{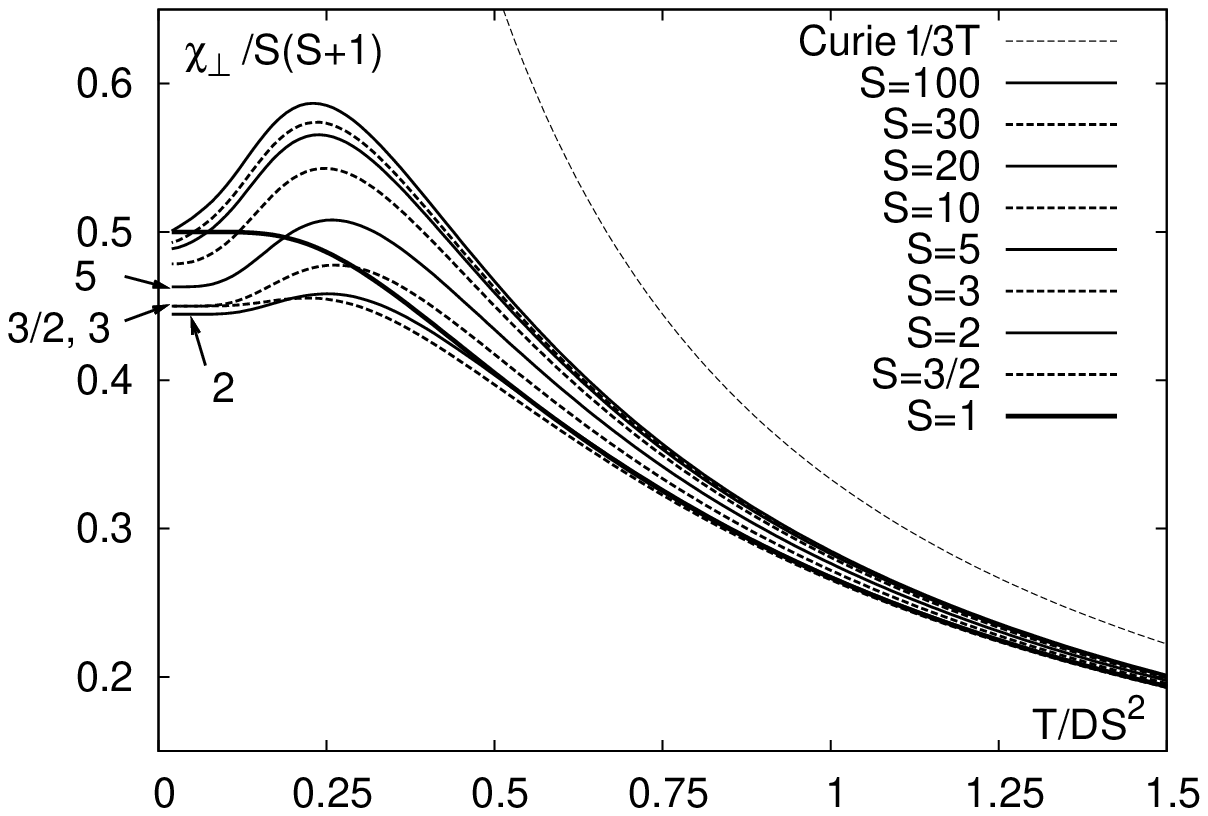}
}
\caption{
Upper panel:
Transverse susceptibility vs.\ $T$ for various $\Sm$, plotted as
$\kT\X/\SSp$.
The two lowest curves correspond to those of $\Sm=1$ and $\Sm=3/2$ in
Fig.~\ref{fig:X:lo-tr-ran}.
In the lower panel we plotted the raw susceptibilities, showing the
peaks developed for $\Sm\geq3/2$, and how they evolve toward the
classical curve.
}
\label{fig:X:manyS:tr}
\end{figure}

The torque picture holds at $T=0$ (Stoner-Wohlfarth \cite{chikazumi}),
and we actually saw that, down there, $\Xtr$ tends to some constant
($\Xtr=\K$ for $\Sm=1$ and $\Xtr=3/4\K$ for $\Sm=3/2$;
Fig.~\ref{fig:X:lo-tr-ran}).
On the other hand, at high $\kT\gg\K$ the anisotropy potential becomes
irrelevant and one should regain $\Xtr=\third\beta\SSp$ once again.
If we insist in a $\kT\X$ plot (Fig.~\ref{fig:X:manyS:tr}), this
quantity will evolve from $\third\SSp$ at high $T$ down to $0$, with
an initial linear slope if $\Xtr(T)$ is nearly constant at low
temperature.

This behavior is what we found before for $\Sm=1$ and $\Sm=3/2$, i.e.,
for the 3-level and 4-level systems.
But $\Sm=3/2$ also exhibited a small and broad maximum in $\Xtr(T)$,
which is absent in $\Sm=1$.
Is this a parity, integer/half-integer effect? or the absence of peak
is just an oddity of $\Sm=1$ and it persists $\forall\,\Sm>1$.
The latter is suggested by the classical model having a maximum too.
And indeed, plotting the bare $\Xtr$ vs.\ $T$ (lower panel of
Fig.~\ref{fig:X:manyS:tr}) we see that the peak is there for all
$\Sm>1$, becoming more apparent for large $\Sm$. %
\footnote{
The initial decrease, and then increase in $\Xtr(T=0)$, is due to our
normalization and scalings (Fig.~\ref{fig:X:manyS:tr}, bottom).
For fixed $\K$, the bare $\Xtr(S,T=0)=1/[2\K(1-1/2\Sm)]$
[Eq.~(\ref{Xtorque})] decreases monotonically with $\Sm$:
$\Xtr(1)=1/\K$, $\Xtr(3/2)=3/4\K$, $\Xtr(2)=2/3\K$, to get halved as
$\Xtr(\Sm\to\infty)=1/2\K$.
But the scaling $\K=1/\Sm^{2}$, plus $\Xtr/\SSp$ gives
$\Xtr\to\Sm^{2}/(2\Sm^{2}+\Sm-1)$, which starts from $1/2$ at $\Sm=1$,
decreases sharply to the minimum $4/9$ at $\Sm=2$, and slowly returns
to $1/2$ as $\Sm\to\infty$.
Notice $\Xtr(3/2)=\Xtr(3)=0.45$ in these units.
} 

Some interpretation can be provided combining the torque picture above
with thermal activation.
A small $T\neq0$ can assist in leaving the potential minima and, on
average, may help reorienting $\J$ toward the transverse field
(increasing the response).
Too high a $T$, however, and the custom thermal misalignment would set
in, decreasing the susceptibility.
Then a peak in between seems natural from the competition of both
processes.

Well, but why is there no peak for $\Sm=1$?
We may answer that, lacking intermediate levels, the $T$ that assists
rotation toward $\m=0$ becomes too large, and we only find the
thermal decrease.
%
%
Quantum mechanically it can be put in the following way.
A transverse field $\bd_{x}$ mixes the states $|\m\rangle$ and
$|\m\pm1\rangle$, and their contribution to $\llangle\Sx\rrangle$
involves the factor
$|\langle \m+1|\Sx|\m\rangle|^{2}/(\el_{\m+1}-\el_{\m})
\times
(\e^{-\beta\el_{\m+1}}-\e^{-\bEm})$.
Then the contribution of $\m=\Sm-1$ is larger than that of $\m=\Sm$,
as the above factor increases when $\m$ climbs the ladder out of
$\m=\Sm$ (mostly because the levels get closer if $\el_{\m}=-\K\m^{2}$
so that $1/(\el_{\m+1}-\el_{\m})$ becomes larger).
However, for $\Sm=1$, the level $\m=0$ has no level above it providing
such an increase.
%
%
Then populating $\m=0$ thermally does not increase $\Xtr$, and there is
no peak for spin one.

As for the behavior of the peaks with $\Sm$, we see that they move
only a little in the scaled units $\kT/\K\,\Sm^{2}$.
This rules out a characteristic temperature of the type
$T_{0}\sim\K(2\Sm-1)$ (as invoked in the longitudinal case), because
then $T_{0}/\K\,\Sm^{2}\sim1/\Sm$, and the peak would shift left and
disappear classically.
We need another governing energy scale; the total barrier
$\dU=\K\,\Sm^{2}$ would do, since is constant in our units.
Indeed, $\exp(-\dU/T_{0})\not\simeq0$ would mark the onset of
misalignment, as the equatorial levels then start to be populated.
Using again the rule-of-thumb $\exp(-x)\simeq0$ at $x\sim4$--$6$, we
would have $\dU/T_{0}\sim4$--$6$.
Then $T_{0}/\K\,\Sm^{2}\sim0.16$--$0.25$, compatible with the peaks'
location in Fig.~\ref{fig:X:manyS:tr}.
\footnote{
In the classical model the peak can be found plotting
$\Xtr=\beta\mu^{2}(R-R')/2R$ \cite{gar2000}, and is located around
$T_{0}\simeq0.23$ \cite{ponomarenko}.
On the other hand, the effective eigenvalue method \cite{cofkalmas93}
gives an effective relaxation time $\tau_{\rm ef}$ showing a similar
maximum \cite[p.~126]{gar2000}.
The reason seems to be that $\tau_{\rm ef}$ can be expressed in terms
of equilibrium averages, and happens to be proportional to
$1-\langle z^{2}\rangle\propto\Xtr$.
} 


\subsection{
response for randomly oriented axes
}
\label{Xran:S}

We conclude with the susceptibility of the ensemble with anisotropy
axes distributed at random.
This case is of experimental interest in powdered samples and liquids
(ferro-fluids~\cite{panpol93}), as well as in solid systems with
orientational disorders.
\begin{figure}[!t]
\centerline{
\includegraphics[width=10.cm]{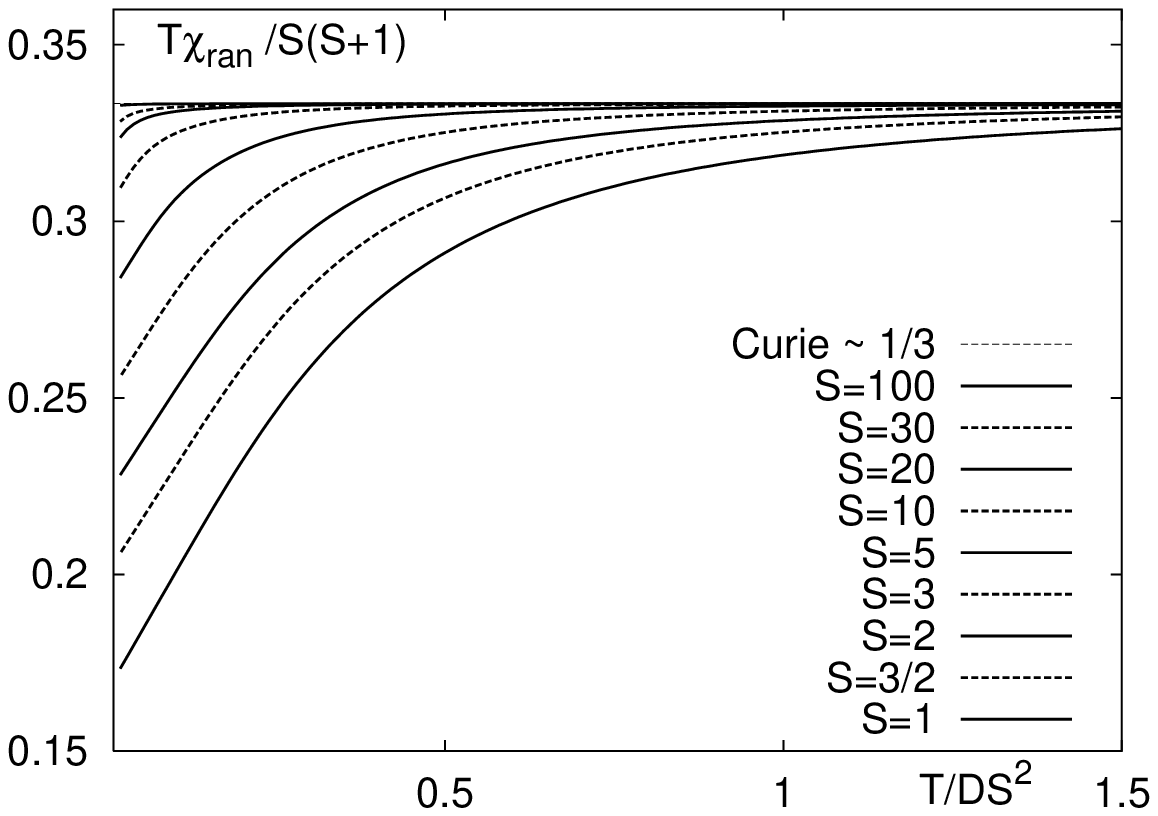}
}
\centerline{
\includegraphics[width=10.cm]{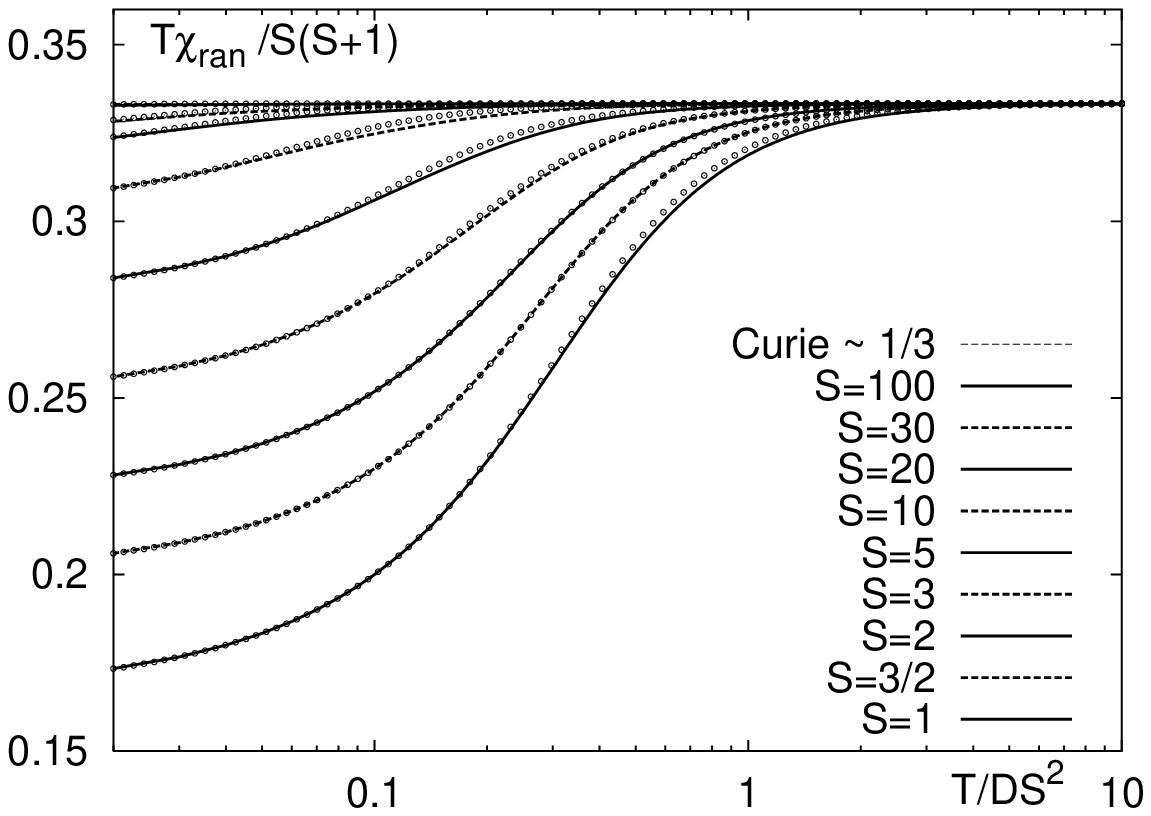}
}
\caption{
Random axes susceptibility vs.\ temperature for various $\Sm$,
presented as $\kT\X/\SSp$.
(Notice the factor $1/2$ difference for $\Sm=1$ with respect to
Fig.~\ref{fig:X:lo-tr-ran}, due to the normalization of $\X$.)
The lower panel shows the same curves but over a wider temperature
range in logarithmic scale, to test the approximate
formula~(\ref{Xran:approx}) (small circles).
}
\label{fig:X:manyS:ran}
\end{figure}


\subsubsection{
restoring the isotropy and the classical limit
}

In classical superparamagnets the orientational average
$\overline{\blo^{2}}=1/3$ and $\overline{\btr^{2}}=2/3$ leads to a
full restoration of isotropy \cite[Fig.~1]{bealiv59} \cite{kru79}
%
\begin{equation}
\label{Xrandomized}
\left.
\begin{array}{rcl}
\Xlo
&=&
\beta\mu^{2}\,
R'/R
\\[0.ex]
\Xtr
&=&
\beta\mu^{2}\,
(R-R')/2R
\end{array}
\right\}
\quad
\leadsto
\quad
\Xran
=
\tfrac{1}{3}
\Xlo
+
\tfrac{2}{3}
\Xtr
%
%
=
\third
\beta\mu^{2}
\;.
\end{equation}
That is 
$R(\alpha)=\int_{0}^{1}\!\drm z\,\exp(\alpha z^{2})$, 
with $\alpha=\dU/T$, and its $\alpha$-derivative $R'$ disappear from
$\Xran$.
This washing away of any trace of the anisotropy constants is not
specific of the uniaxial model $\Hs=-\K\,\Sz^{2}$, but it holds for
any classical $\Hs$ with the reflection symmetry $\Hs(-\J)=\Hs(\J)$
(see~\cite{garjonsve2000} and references therein).

Quantum mechanically, however, we know that this cannot be exact.
In Fig.~\ref{fig:X:lo-tr-ran} we showed that for $\Sm=1$ and $\Sm=3/2$
the random axes averaging indeed restores isotropy, $\Xran\simeq\Xc$,
over a wider temperature range.
And that this occurs even with $\Xlo$ well on its way to
$\Xising=\beta\Sm^{2}$, but at a certain temperature the
susceptibility starts to deviate noticeably downward (it cannot catch
up with $\third\beta\SSp$).
That temperature is of the order of $\K$ for $\Sm=1$ and $2\K$ for
$\Sm=3/2$ (Fig.~\ref{fig:X:lo-tr-ran}), suggesting that the
relevant/governing energy scale is again $T_{\rm I}=\K(2\Sm-1)$.
Indeed, in units of $\K\,\Sm^{2}$ the extent of the bent range would
be from $T=0$ to $T_{\rm I}/\K\,\Sm^{2}\sim2/\Sm$
(see Fig.~\ref{fig:X:manyS:ran}).
This smoothly gives the connection with the classical result of full
restoration of isotropy by letting $1/\Sm\to0$.


\subsubsection{
approximate formula for $\Xran$
}

To confirm the previous estimates and scalings we have derived an
approximate formula for $\Xran$ based on a few-levels treatment.
It may also provide some insight in the origin of the deviations
from $\Xc=\third\beta\SSp$.
\begin{table}[!tb]
\begin{center}
\begin{tabular}{c|cccl}
$\m$
&
$-\bEm$
&
$\lf_{\m}^{2}$
&
$\beta\tf_{\m\,\m+1}$
&
\\
\hline
$-\Sm$
&
$\ds\,\Sm^{2}$
&
$2\Sm$
&
$-\ds(2\Sm-1)$
&
$\bydefr-\Omega$
\\
$-\Sm+1$
&
$\ds\,(\Sm-1)^{2}$
&
---
&
---
&
\\
$+\Sm-1$
&
$\ds\,(\Sm-1)^{2}$
&
$2\Sm$
&
$+\ds(2\Sm-1)$
&
$\bydefr+\Omega$
\\
$+\Sm$
&
$\ds\,\Sm^{2}$
&
$0$
&
---
&
\\
\end{tabular}
\end{center}
\caption{
Lowest energy levels for spin $\Sm$ at zero field
$-\bEm=\ds\,\m^{2}$,
ladder factors $\lf_{\m}^{2}=\SSp-\m(\m+1)$,
and the transition frequencies
$\beta\tf_{\m\,\m+1}=\beta(\el_{\m}-\el_{\m+1})=\ds(2\m+1)$.
The approximate partition function is
$Z_{0}
\simeq
2(\e^{-\beta\el_{\Sm}}+\e^{-\beta\el_{\Sm-1}})
=
2\,\e^{-\beta\el_{\Sm}}(1+\e^{-\Omega})$
(cf.~\cite[App.~A]{lopetal2005}).
}
\label{table:S}
\end{table}

We started looking at Eq.~(\ref{Xtr}) for $\Xtr$.
One notices that letting $K_{\m}\to1$ (for instance, considering
$K(\beta\tf)$ plus $\tf\sim1/\Sm$), the susceptibility is left as
$\Xtr^{\rm iso}=(\beta/2)\llangle\lf_{\m}^{2}\rrangle$,
with
$\llangle\lf_{\m}^{2}\rrangle
\bydefl
\sum_{\m}
\lf_{\m}^{2}
\,
\e^{-\bEm}/\Z_{0}$.
Then, combining the average of the ladder factor
$\lf_{\m}^{2}=\SSp-\m(\m+1)$ with the longitudinal susceptibility
gives
$\tfrac{1}{3}\Xlo+\tfrac{2}{3}\Xtr^{\rm iso}=\third\beta\SSp$,
explaining the notation $\Xtr^{\rm iso}$.
The actual susceptibility is obtained by adding and subtracting
$\llangle\lf_{\m}^{2}\rrangle$ to Eq.~(\ref{Xtr})
%
\begin{equation}
\label{Xtr:rewritten}
\Xtr
=
\Xtr^{\rm iso}
+
(\beta/2)\llangle\lf_{\m}^{2}(K_{\m}-1)\rrangle
\;.
\end{equation}
This form, times $2/3$ and combined with the longitudinal
part~(\ref{Xlo}), gives
$\Xran=\Xc+(\beta/3)\llangle\lf_{\m}^{2}(K_{\m}-1)\rrangle$.
Thus in the second term (of transverse origin) we have isolated the
source of deviation from isotropy, but we have not done approximations
yet.%
\footnote{ 
By analogy with the classical superparamagnetic
susceptibilities~(\ref{Xrandomized}) one could introduce
$\MMz=:\SSp\,R'/R$ (which defines $R'/R$), and write
($\bar\kT=\kT/\SSp$)
%
\[
\bar\kT\,
\Xlo
=
R'/R
\qquad
\bar\kT\,
\Xtr
=
(R-R')/2R
+
{\rm corrections}
\]
The first term in $\Xtr$ corresponds to $\Xtr^{\rm iso}$ (i.e., when
combined with $\Xlo$ and randomized gives Curie), and the corrections
correspond to
$\llangle\lf_{\m}^{2}(K_{\m}-1)\rrangle$.
} 

To compute the suspect term $\llangle\lf_{\m}^{2}(K_{\m}-1)\rrangle$
we just include the contribution of the two lowest levels (at each
side; see Table~\ref{table:S}), getting the following susceptibility
%
\begin{equation}
\label{Xran:approx}
\Xran
\simeq
\frac{1}{3}
\underbrace{
\beta\Sm^{2}
}_{}
+
\frac{2}{3}
\underbrace{
\frac{1}{2\K(1-1/2\Sm)}
}_{}
\frac{1-\e^{-\Omega}}{1+\e^{-\Omega}}
\;,
\qquad
\Omega=\beta\K(2\Sm-1)
\;.
\end{equation}
From this expression we identify the Ising $\Xlo\simeq\beta\Sm^{2}$ at
low temperatures.
Besides, invoking $\Omega\bydefl\beta\K(2\Sm-1)\gg1$ at low $T$, one
can read off the torque transverse susceptibility $\forall\,\Sm$,
namely
%
\begin{equation}
\label{Xtorque}
\Xtr(T=0)
=
\frac{1}{2\K(1-1/2\Sm)}
\;.
\end{equation}
%
%
This gives the familiar values $\Xtr=1/\K$ for $\Sm=1$ and
$\Xtr=3/4\K$ for $\Sm=3/2$.

Equation~(\ref{Xran:approx}) is governed by the level difference
$\Omega\bydefl\beta\tf_{\Sm,\Sm-1}$, as expected from our lowest-level
approximation.
We have plotted it together with the exact $\Xran$ curves in the lower
panel of Fig.~\ref{fig:X:manyS:ran}.
The description it provides is reasonably good at all temperatures,
for all spin values, and quite good for $\Sm=3/2$ and $2$.

Multiplying now Eq.~(\ref{Xran:approx}) across by $\kT$ and using
$\beta\K(1-1/2\Sm)=\Omega/2\Sm$ we can write the compact form
%
\begin{equation}
\label{Xran:approx:compact}
\kT\,\Xran
\simeq
\frac{\Sm}{3}
\bigg(
S
+
\frac{2}{\Omega}
\,
\frac{1-\e^{-\Omega}}{1+\e^{-\Omega}}
\bigg)
\;,
\qquad
\Omega=\beta\K(2\Sm-1)
\;.
\end{equation}
This gives the $\Sm$-dependent intercept of the $\X$ axis at $T=0$ of
Fig.~\ref{fig:X:manyS:ran} and the initial linear growth at low
temperatures ($1/\Omega\propto T$ with $\e^{-\Omega}\simeq0$).
But Eq.~(\ref{Xran:approx:compact}) happens to capture as well the
restoring of Curie at high $T$ ($\Omega\ll1$).
Indeed, Taylor expanding 
$(2/\Omega)
(1-\e^{-\Omega})/(1+\e^{-\Omega})
\simeq
1-\Omega^{2}/12$,
the constant term nicely produces $\third\SSp$, the Curie constant.
The second term can then be used to estimate the onset of deviations
from $\Xc$.
Fixing a $5$\,\% deviation, i.e., $\Omega^{2}/12\simeq0.05$, and
writing $\Omega=(\K\,\Sm^{2}/\kT)(2\Sm-1)/\Sm^{2}$, yields
$\kT/\K\,\Sm^{2}\sim5/2\Sm$, in agreement with our previous estimate
$2/\Sm$.

As for the use of the approximate $\Xran$ one should bear in mind that
Eq.~(\ref{Xran:approx}) was derived under low $T$ conditions (few
populated levels).
However, the reasonable agreement with the exact curves in the whole
temperature range, and for all $\Sm$, indicates that it could be used
safely in modelization of uniaxial magnets with axes distributed at
random.


\section{
summary
}

The understanding of the properties of paramagnets belongs to a long
tradition linking magnetism, quantum mechanics and statistical
mechanics.
Our aim here has been to extend the theoretical framework to permit
the study of some equilibrium problems for arbitrary values of the
spin.
In this frame, one can connect from landmark results for quantum
paramagnets (Curie-Brillouin, transverse response of anisotropic
spins) all the way up to the theory of classical superparamagnets,
developed for magnetic nanoparticles, and revived with the young
molecular magnetic clusters (single-molecule magnets).
We have used the language of magnetism throughout, but the formalism
is closely related with the effective big-spin description of
collections of 2-level systems used in atom optics and two-mode Bose
condensates \cite{ACGT72,miletal97,angvar2001}.

We focused on uniaxial spins and the temperature dependence of the
magnetic susceptibility, due to its traditional significance and its
routine use as characterization tool.
We investigated three features:
(1) the crossover of the longitudinal susceptibility from Curie to the
    2-state regime, induced by the magnetic anisotropy,
(2) the peak in the transverse $\Xtr$ vs.\ $T$, and
(3) deviations of the random ensemble susceptibility from the Curie
    law (absent in the classical limit).
We identified and characterized the relevant energy/temperature
parameters governing the phenomenology, and studied how they scale with
$\Sm$.
We did this starting from small spins $\Sm=1$,~$3/2$,~$2$, then moderate
$\Sm=5$,~$10$,~\dots and eventually big spins $\Sm=30$,~$50$,~$100$,
connecting with classical superparamagnetic phenomenology.

The equations we employed do not rely on Van Vleck's method, as they
follow from the general Kubo correlator formalism of linear-response
theory.
We worked and particularized this formalism, with the above problems
in mind, to produce ready-to-use formulas [Eqs.~(\ref{Xlo})
and~(\ref{Xtr})], which only require the input of the unperturbed
spectrum and angular-momentum ladder factors.

We also derived approximate expressions and assessed their ranges of
validity.
This turned out to be quite good for the lowest-levels
approximation~(\ref{Xran:approx}) to the susceptibility of the
randomly-oriented ensemble.
It could be safely used as a compact modelization of susceptibility
data for arbitrary $\Sm$ in such an experimentally relevant case.
In real magnets, however, the unperturbed Hamiltonian will include
terms non-diagonal in the standard basis, like
$\K\,\Sm_{\pm}^{2}$ or $K\Sm_{\pm}^{4}$.
These terms were not accounted for in the derivation of the
approximate expression~(\ref{Xran:approx}) and, when relevant, one
should turn back to the more general equation~(\ref{X:unbiased}).


\section*{Acknowledgments}

In this work J.L.G.P. and J.G. were supported by 
"NUS YIA, WBS grant {No.~R-144-000-195-101}" and F.L. by project
 NABISUP (DGA).







\newpage\tableofcontents
\enlargethispage{1.cm}

\end{document}